\pdfoutput=1

\documentclass{aastex6}

\begin{document}

\title{Activity Analyses for Solar-type Stars Observed with \emph{Kepler}. II. Magnetic Feature Versus Flare Activity}
\author{Han He\altaffilmark{1,2}, Huaning Wang\altaffilmark{1,2}, Mei Zhang\altaffilmark{1,2}, Ahmad Mehrabi\altaffilmark{3,4}, Yan Yan\altaffilmark{1}, and Duo Yun\altaffilmark{1}}
\altaffiltext{1}{CAS Key Laboratory of Solar Activity, National Astronomical Observatories, Chinese Academy of Sciences, Beijing, China; hehan@nao.cas.cn}
\altaffiltext{2}{School of Astronomy and Space Science, University of Chinese Academy of Sciences, Beijing, China}
\altaffiltext{3}{Department of Physics, Bu Ali Sina University, 65178, 016016, Hamedan, Iran}
\altaffiltext{4}{School of Astronomy, Institute for Research in Fundamental Sciences (IPM), 19395-5531, Tehran, Iran}

\begin{abstract}
  The light curves of solar-type stars present both periodic fluctuation and flare spikes. The gradual periodic fluctuation is interpreted as the rotational modulation of magnetic features on the stellar surface and is used to deduce magnetic feature activity properties. The flare spikes in light curves are used to derive flare activity properties. In this paper, we analyze the light curve data of three solar-type stars (KIC 6034120, KIC 3118883, and KIC 10528093) observed with \emph{Kepler} space telescope and investigate the relationship between their magnetic feature activities and flare activities. The analysis shows that: (1) both the magnetic feature activity and the flare activity exhibit long-term variations as the Sun does; (2) unlike the Sun, the long-term variations of magnetic feature activity and flare activity are not in phase with each other; (3) the analysis of star KIC 6034120 suggests that the long-term variations of magnetic feature activity and flare activity have a similar cycle length. Our analysis and results indicate that the magnetic features that dominate rotational modulation and the flares possibly have different source regions, although they may be influenced by the magnetic field generated through a same dynamo process.
\end{abstract}


\section{Introduction} \label{sec:intro}

High-precision and continuous light curves of solar-type stars observed with \emph{Kepler} space telescope \citep{2010Sci...327..977B, 2010ApJ...713L..79K, 2016RPPh...79c6901B} and other space missions like \emph{CoRoT} \citep{2009A&A...506..411A} and \emph{MOST} \citep{2003PASP..115.1023W} present two significant ingredients: periodic fluctuation (rotational modulation) and flare spikes. The rotational modulation on the stellar brightness variation is interpreted as being caused by the corotating magnetic features (e.g., dark starspots and bright faculae) on the stellar surface \citep{1981ApJ...250..276V, 1983ApJ...275..752B, 2011A&A...529A..89D, 2014ApJS..211...24M, 2015ApJS..221...18H, 2015MNRAS.452.2745S, 2016A&A...586A..14H, 2016A&A...589A..46S, 2017ApJ...834..207M}, and the sudden spikes in the light curves is explained as flare phenomenon originated from the starspot regions associated with intense and concentrated magnetic field \citep{2011AJ....141...50W, 2012Natur.485..478M, 2013ApJS..207...15K, 2013ApJ...771..127N, 2013ApJS..209....5S, 2014ApJ...797..121H, 2015MNRAS.447.2714B, 2016ApJ...829...23D, 2016NatCo...711058K, 2016AcASn..57....9Y, 2017ChA&A..41...32D, 2017ApJS..232...26V}. In this paper, we analyze the relationship between these two aspects of stellar activity (for a general review of stellar activity, please refer to, e.g., \citealt{2017SCPMA..6019601C}), that is, the magnetic feature activity properties manifested by the fluctuation characteristics of light curves versus the flare activity properties manifested by the flare spikes in light curves, for three solar-type stars observed with \emph{Kepler}.

\citet{2015ApJS..221...18H} (Paper I of this series) suggested two quantitative measures, $i_{\rm AC}$ and $R_{\rm eff}$, as the magnetic activity proxies of solar-type stars based on the \emph{Kepler} light curve observations. The first proxy $i_{\rm AC}$ (autocorrelation index) describes the degree of periodicity of a light curve, which indicates the stability of the magnetic features that dominate the rotational modulation. The second proxy $R_{\rm eff}$ measures the effective fluctuation range of a light curve, which indicates the spatial size or coverage of magnetic features and hence the intensity of magnetic activity \citep{2010ApJ...713L.155B, 2011AJ....141...20B, 2013ApJ...769...37B, 2010Sci...329.1032G, 2011ApJ...732L...5C, 2012A&A...539A.137M, 2013MNRAS.432.1203M, 2017A&A...603A..52R}. The analysis by \citet{2015ApJS..221...18H} illustrates that both the two proxies can reflect the cyclic variation of magnetic activities of solar-type stars. Their result shows that, for a solar-type star, the time variations of the two magnetic proxies may be in the same phase (positive correlation) or in the opposite phase (negative correlation), which implies two different magnetic activity behaviors of solar-type stars. As demonstrated by \citet{2015ApJS..221...18H} using the solar light curve data \citep{1997SoPh..175..267F} obtained by the \emph{SOHO} spacecraft \citep{1995SoPh..162....1D}, the Sun is a negative correlation star. \citet{2017ApJ...834..207M} further analyzed the correlation between the two proxies for a large sample of G-type main sequence \emph{Kepler} targets. They found that the number of positive correlation stars is much larger than the number of negative correlation stars in \emph{Kepler} sample, and the positive correlation stars tend to have shorter rotation period and larger magnitude of light curve variation than the negative correlation stars.

On the Sun, major flares occur in the sunspot regions holding a complex magnetic geometry \citep[e.g.,][]{1996ApJ...456..861W, 2015SCPMA..58.5682W, 2002A&ARv..10..313P, 2003ApJ...595.1296L, 2007ApJ...656.1173L, 2008LRSP....5....1B, 2008AdSpR..42.1450H, 2014JGRA..119.3286H, 2009RAA.....9..687W, 2011LRSP....8....6S, 2014ApJ...786...72Y, 2017ApJ...834..150Y}, and the sunspot number presents a 11 yr solar cycle \citep[e.g.,][]{2010LRSP....7....6P, 2015LRSP...12....4H}. Statistical studies \citep[e.g.,][]{2012ApJ...754..112A, 2015LRSP...12....4H} show that the occurrence frequency of solar flares has a clear cyclic variation in pace with the solar cycle, that is, the flare frequency is higher around solar maxima and is lower at solar minima. (Since it is sunspots that dominate the rotational modulation in the light curves of the Sun, the 11 yr solar cycle of sunspot number can also be reflected by the cyclic variation of the fluctuation range of solar light curves, see the quantitative analysis using the measure $R_{\rm eff}$ by \citealt{2015ApJS..221...18H}.) On the other hand, \citet{2012ApJ...754..112A} found that the power-law distribution of the magnitude of solar flares (defined as the background-subtracted soft X-ray peak flux) is invariant throughout the solar cycle. This means that even during a cycle minimum the high magnitude solar flares can also occur, though with very low frequency \citep[see also][]{2015LRSP...12....4H}. In short, the sunspot number (indicating sunspot coverage on the solar disk) affects the occurrence frequency of solar flares but not the distribution of flare magnitude.

For solar-type stars, the relations between stellar flare activity and magnetic activity have been studied by many authors in the literature. The \emph{Kepler} dataset possesses a large sample of stellar flares and was extensively used in these studies. For example, in \citet{2013ApJS..209....5S}, the authors extended the pioneer work of \citet{2012Natur.485..478M} and compiled a superflare catalog of solar-type (G-type) stars based on the observational data of \emph{Kepler}. By utilizing this catalog, \citet{2013ApJ...771..127N} found that the energy of superflares is related to the amplitude of brightness variation of the host stars, that is, the stars with higher brightness variation amplitude tend to have more energetic flares. Similar results (flare magnitude vs. amplitude of stellar brightness variation) were also obtained for other kinds of stars with different spectral types or effective temperatures \citep{2011AJ....141...50W, 2015MNRAS.447.2714B}, but the stellar brightness variability is only weakly correlated with the flare occurrence frequency \citep{2011AJ....141...50W}. \citet{2012Natur.485..478M} reported a relation between the flare occurrence frequency and the stellar brightness variation amplitude for solar-type stars in the supplementary information of their paper, but this relation is not confirmed in the follow-up studies using a larger sample of superflare stars \citep{2013ApJ...771..127N, 2013ApJS..209....5S}. The amplitude of stellar brightness variation is commonly attributed to the coverage of starspots, yet the statistical result with stellar flare frequency implies a contradiction from the result for the Sun mentioned above.

The stellar magnetic activity can also be reflected by chromospheric observations \citep{1913ApJ....38..292E, 1978ApJ...226..379W, 2008LRSP....5....2H}. By employing a chromospheric activity index \citep{1978PASP...90..267V} derived from the spectrum observations \citep{2015ApJS..220...19D} of the Ca \textsc{ii} H and K lines for \emph{Kepler} objects using the ground-based Large Sky Area Multi-Object Fiber Spectroscopic Telescope \citep[LAMOST;][]{2012RAA....12.1197C}, \citet{2016NatCo...711058K} demonstrated that the superflare stars in the catalog of \citet{2013ApJS..209....5S} generally possess a higher chromospheric activity level than average stars, including the Sun, and this higher activity level of superflare stars is consistent with the result of the stellar spectrum analysis performed by \citet{2015PASJ...67...33N}. \citet{2016NatCo...711058K} also found that their chromospheric activity index values are positively correlated with the stellar periodic photometric variability amplitudes measured by \citet{2014ApJS..211...24M} using the \emph{Kepler} data down to 1000 parts per million (ppm), which is just the order of amplitude of the Sun. (Note that the concept of stellar brightness variation amplitude or periodic photometric variability amplitude is equivalent to the light curve fluctuation range measure $R_{\rm eff}$ employed in this work.) The stellar photometric variability amplitude is attributed to the coverage of starspots as usual, yet the emission of Ca \textsc{ii} H and K lines employed to derive the chromospheric activity index mainly comes from the plage regions in chromosphere \citep[e.g.,][]{1975ApJ...200..747S, 1989ApJ...337..964S}. Chromospheric plages correspond to enhanced network magnetic field and facula regions in the photosphere \citep{2006RPPh...69..563S}, which might surround but are not necessarily associated with starspots \citep{1975ApJ...200..747S, 1989ApJ...337..964S}.

In previous studies, relations between stellar magnetic activity and flare activity for solar-type stars were analyzed mainly based on large samples of stars, and lacked time evolution information. In this work, we perform the analysis of magnetic feature activity versus flare activity for three individual solar-type stars. Like the approaches for the Sun, we investigate the time variations of stellar flare activity and magnetic activity, with a particular interest in their phase relationships. The investigated superflare stars are selected from the catalog of \citet{2013ApJS..209....5S}. These selected \emph{Kepler} stars possess different rotation periods and have different flare frequencies, which are described in Section \ref{sec:objects}. For each star, two time-series components are extracted from the original \emph{Kepler} light curves: the rotational modulation component (caused by corotating magnetic features) and the flare component (constituted by flare spikes). The procedure and products of the \emph{Kepler} light curve data processing are described in Section \ref{sec:data}. The magnetic feature activity properties of the stars are evaluated based on the rotational modulation component using the two magnetic proxies introduced in Paper I \citep{2015ApJS..221...18H}. The flare activity properties are derived from the flare component via several flare indexes. The methods for the magnetic proxies and flare index evaluations are described in Section \ref{sec:parameters}. The relationships between the magnetic feature activity and the flare activity are analyzed in Section \ref{sec:relations}. Section \ref{sec:sum} presents a summary and discussion.

\section{Superflare Stars Investigated} \label{sec:objects}

The individual \emph{Kepler} stars analyzed in this work are selected from the superflare star catalog of \citet{2013ApJS..209....5S}. For the aim of studying time evolution of stellar activity properties, the stars should have continuously observed light curve data, i.e., long-cadence data \citep{2010ApJ...713L.120J} from quarters 2 to 16 (Q2--Q16). The data in Q0, Q1, and Q17 are not included in this work because the three quarters are too short (see illustration in Paper I) to yield a compatible result with the other full-length quarters \citep[about three months;][]{2010ApJ...713L.115H}. We picked out all the stars with full long-cadence data in Q2--Q16 from the catalog of \citet{2013ApJS..209....5S} and sorted them by their flare-counting numbers given by \citet{2013ApJS..209....5S}. The star at the top of the list (i.e., with the highest flare occurrence frequency) is KIC 6034120. In this paper, our analyses of stellar magnetic activity versus flare activity are mainly based on this flare-abundant star (notice that the same \emph{Kepler} object has been quoted in the paper by \citet{2012Natur.485..478M} as a representative superflare star). For comparison, two other superflare stars, KIC 3118883 and KIC 10528093, with different rotation periods and flare occurrence frequencies, are also employed in this study. The stellar parameters and basic \emph{Kepler} data information (effective temperature, $T_{\rm eff}$; surface gravity, $\log g$; metallicity, [Fe/H]; radius, $R$; \emph{Kepler} magnitude, $Kp$; rotation period, $P_{\rm rot}$; and number of flares identified by \citealt{2013ApJS..209....5S}) of the three investigated superflare stars are given in Table \ref{tab:star-param}.

\floattable
\begin{deluxetable}{lccclccDc}
\tablecaption{Stellar Parameters and Basic \emph{Kepler} Data Information of the Three Investigated Superflare Stars \label{tab:star-param}}
\tablehead{
\emph{Kepler} ID & \colhead{} & \colhead{$T_{\rm eff}$\tablenotemark{a}} & \colhead{$\log g$\tablenotemark{a}} &
\colhead{[Fe/H]\tablenotemark{a}} & \colhead{$R$\tablenotemark{a}} & \colhead{$Kp$\tablenotemark{a}} &
\twocolhead{$P_{\rm rot}$\tablenotemark{b}} & \colhead{Number of Flares Identified}\\
\colhead{} & \colhead{} & \colhead{(K)} & \colhead{(cgs)} &
\colhead{} & \colhead{$(R_\odot)$} & \colhead{} &
\twocolhead{(days)} & \colhead{by \citet{2013ApJS..209....5S}\tablenotemark{c}}
}
\startdata
KIC 6034120  & & 5407 & 4.654 & $-0.290$ & 0.770 & 14.882 &  5.724  & &  45 \\
KIC 3118883  & & 5195 & 4.458 & $-0.328$ & 0.967 & 14.357 &  8.653  & &  17 \\
KIC 10528093 & & 5143 & 4.526 & $-0.546$ & 0.883 & 13.563 & 12.232  & &  18 \\
\enddata
\tablenotetext{a}{The values of effective temperature ($T_{\rm eff}$), surface gravity ($\log g$), metallicity ([Fe/H]), radius ($R$), and \emph{Kepler} magnitude ($Kp$) are taken from the \emph{Kepler} Input Catalog \citep[KIC;][]{2011AJ....142..112B}.}
\tablenotetext{b}{The values of rotation period ($P_{\rm rot}$) of \emph{Kepler} stars are taken from the measurements by \citet{2014ApJS..211...24M}.}
\tablenotetext{c}{Only the \emph{Kepler} data in Q0--Q6 were employed for searching flares by \citet{2013ApJS..209....5S}.}
\end{deluxetable}

\section{Light Curve Data Processing} \label{sec:data}
We adopt the PDC (Pre-search Data Conditioning) flux data \citep{2012PASP..124.1000S, 2012PASP..124..985S, 2014PASP..126..100S} of the long-cadence light curve product in \emph{Kepler} Data Release 25 \citep{KeplerDataRelease25} for our study. The PDC data are the output of the PDC module of the \emph{Kepler} science processing pipeline \citep{2010ApJ...713L..87J, KeplerDataProcessingHandbook} in which the systematic errors in the raw data are corrected \citep{2012PASP..124.1000S, 2012PASP..124..985S, 2014PASP..126..100S}. The time resolution of the long-cadence PDC flux data is one data point every 29.4 minutes \citep{2010ApJ...713L.120J}.

The absolute values of the original \emph{Kepler} PDC flux data feature discontinuities across quarter boundaries owing to the $90^{\circ}$-roll of the telescope and hence the shift of CCD modules between quarters (\citealt{2016RPPh...79c6901B}; \citealt{KeplerDataHandbook}; note that the \emph{Kepler} mission pursues high differential photometric precision instead of absolute flux accuracy). Considering this data property, we process the \emph{Kepler} flux data of each quarter separately, as done in Paper I. Our aim is to extract two time-series components from the original \emph{Kepler} light curves: one is the rotational modulation component (associated with magnetic features); and the other is the flare component (constituted by flare spikes). Then, the relation between stellar magnetic feature activity and flare activity can be analyzed. The procedure and products of the \emph{Kepler} light curve data processing are described in the following subsections.

\subsection{Extracting the Rotational Modulation Component} \label{subsec:f_M}

\begin{figure}
  \epsscale{0.46}
  \plotone{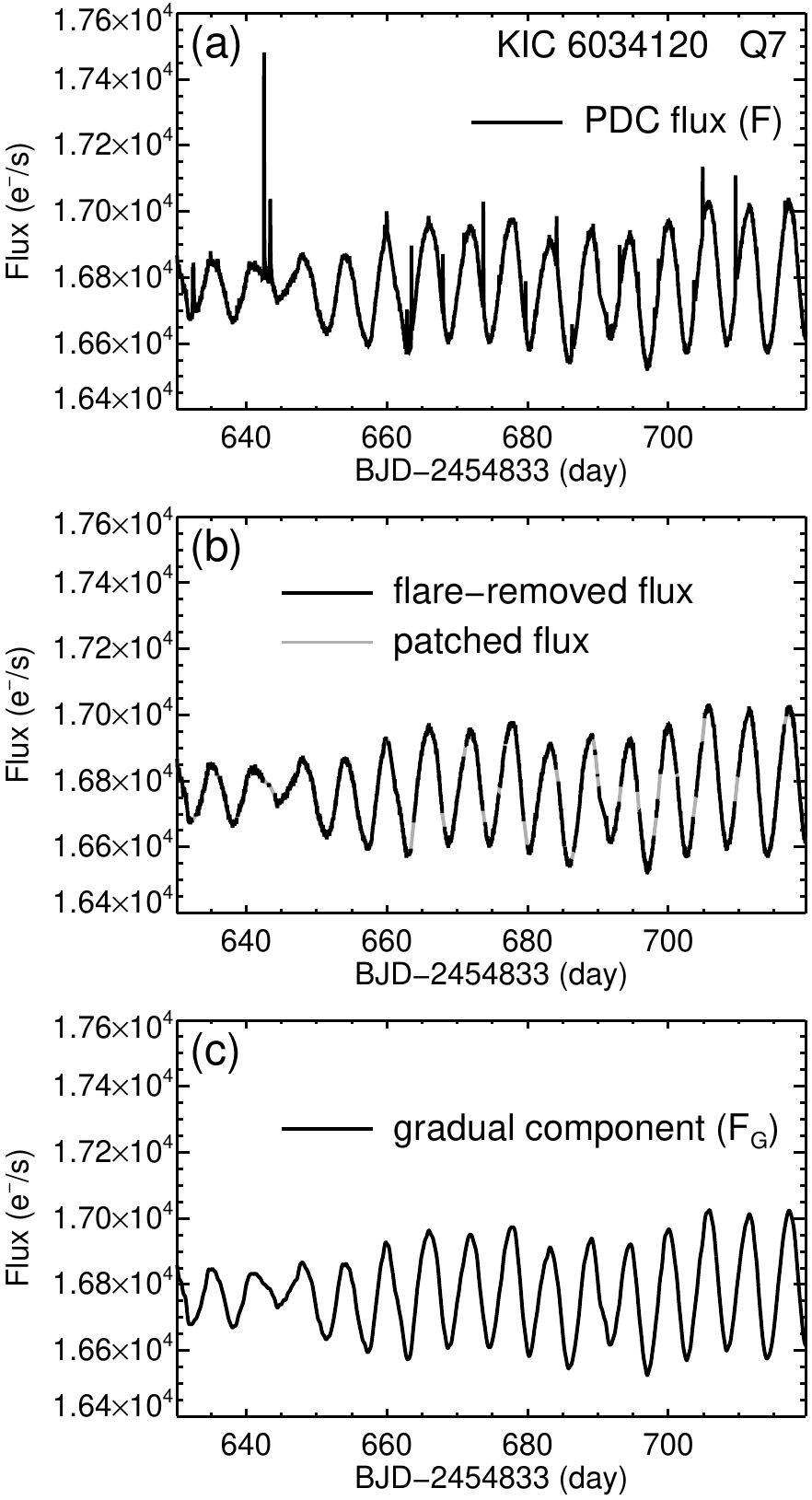}
  \caption{Extraction of the gradual fluctuation component from the observed \emph{Kepler} light curve. (a) PDC flux $F$ (observed light curve). (b) Flare-removed flux curve (black) and patched flux curve (gray). (c) Extracted gradual component flux $F_{\rm G}$. The example data are from Q7 of KIC 6034120. \label{fig1}}
\end{figure}

We use the light curve data of KIC 6034120 in Q7 to demonstrate the procedure of data processing. The first step is to extract a gradual fluctuation component from the observed \emph{Kepler} light curve (PDC flux data), which is illustrated in Figure \ref{fig1}. Figure \ref{fig1}(a) displays the plot of the observed PDC flux (denoted by $F$; note that the missing data points in the original PDC data have been patched up by linear interpolation). To extract the gradual fluctuation component, we identify the prominent flare spikes (with the aid of an automatic software tool) in the original light curve $F$ and then remove all the data points that constitute the identified flare spikes from the light curve. The flare-removed flux curve is shown in Figure \ref{fig1}(b) in black. The void data points left by the flare spikes are subsequently patched up through linear interpolation. The patched flux curve is plotted in Figure \ref{fig1}(b) in gray. As done in Paper I, we filter out the transient variation component (mainly noises, with some minor flares that were not identified in the previous operation) in the patched light curve, and also the stiff transitions at the two ends of the linear interpolations for the flare voids, using a Fourier-based low-pass filter. The upper cutoff frequency of the filter is set as $1/0.5$ day$^{-1}$ which was determined empirically with the consideration that 0.5 day is roughly the upper limit of flare duration time of most stellar flares \citep{2015MNRAS.447.2714B}. The flux curve after filtering is the desired gradual component (denoted by $F_{\rm G}$ where the letter `G' refers to `gradual') of the original \emph{Kepler} light curve. The derived $F_{\rm G}$ curve is plotted in Figure \ref{fig1}(c).

\begin{figure}
  \epsscale{0.46}
  \plotone{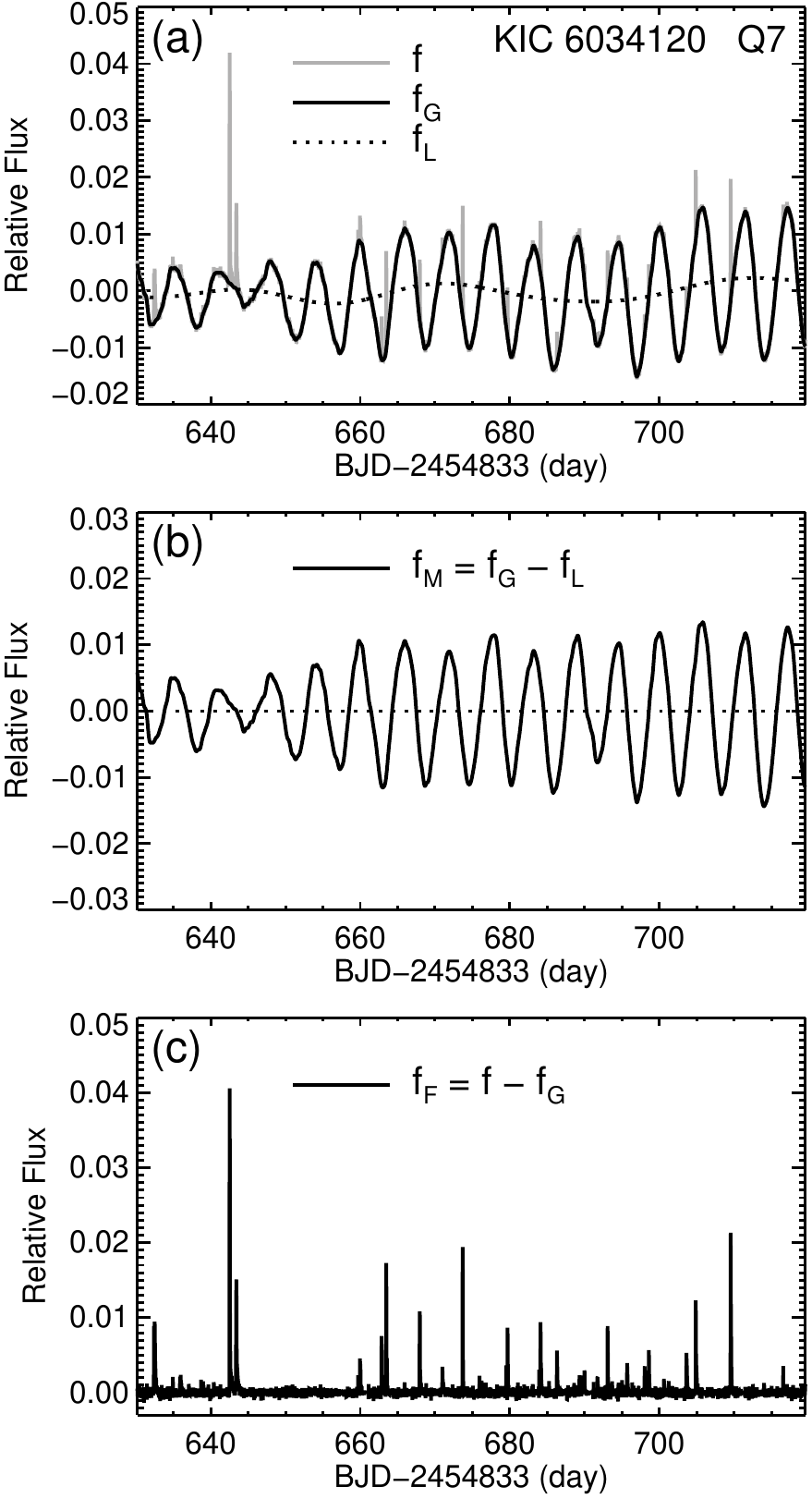}
  \caption{Extraction of rotational modulation component and flare component from the relative flux of the \emph{Kepler} light curve. (a) Total relative flux $f$ (gray solid curve), gradual component $f_{\rm G}$ (black solid curve), and long length-scale trend $f_{\rm L}$ (dotted curve). (Note that the $f_{\rm G}$ curve overlays the $f$ curve.) (b) Extracted rotational modulation component $f_{\rm M}$. (c) Extracted flare component $f_{\rm F}$. The example data are from Q7 of KIC 6034120. \label{fig2}}
\end{figure}

The second step is to calculate the relative fluxes $f$ and $f_{\rm G}$ based on $F$ and $F_{\rm G}$, respectively, to dispel the discontinuity issue at quarter boundaries since all the relative fluxes of different quarters fluctuate around zero (see Paper I). The baseline value $F_0$ for deriving the relative fluxes in the given quarter is adopted as
\begin{equation}
 F_0={\rm median}(F_{\rm G}).
\end{equation}
Then, the total relative fluxes $f$ and the gradual component $f_{\rm G}$ can be calculated by the following equations:
\begin{equation}
 f=\frac{F-F_0}{F_0},
\end{equation}
\begin{equation}
 f_{\rm G}=\frac{F_{\rm G}-F_0}{F_0}.
\end{equation}
The derived $f$ and $f_{\rm G}$ flux curves of the example data are shown in Figure \ref{fig2}(a) (in gray and black, respectively).

The third step is to extract the pure rotational modulation components from the $f_{\rm G}$ flux. As shown in Figure \ref{fig2}(a), some residual, long length-scale trend remained in $f_{\rm G}$. We filter out this long length-scale trend (denoted by $f_{\rm L}$ where the letter `L' refers to `long length-scale'; see the dotted curve in Figure \ref{fig2}(a)) through a high-pass filter acting on the $f_{\rm G}$ flux data. The lower cutoff frequency of the high-pass filter is set as $1/1024$ long-cadence$^{-1}$ (about $1/21$ day$^{-1}$; note the one long-cadence $\approx$ 29.4 minutes), as adopted by the multiscale MAP (maximum a posteriori) algorithm of the \emph{Kepler} science processing pipeline \citep{2014PASP..126..100S}. The resulting flux is the expected rotational modulation component of the original \emph{Kepler} light curve (in relative flux expression) and is denoted by $f_{\rm M}$ (where the letter `M' refers to `modulation'). In short,
\begin{equation}
 f_{\rm M}=f_{\rm G}-f_{\rm L}.
\end{equation}
The $f_{\rm M}$ flux curve of the example data is shown in Figure \ref{fig2}(b). The analysis of magnetic feature activities in this work is based on the $f_{\rm M}$ flux data.

\subsection{Extracting the Flare Component} \label{subsec:f_F}
The flare component of the original \emph{Kepler} light curve in a relative flux expression (denoted by $f_{\rm F}$, where the letter `F' refers to `flare') can be obtained by subtracting the gradual component $f_{\rm G}$ from $f$, that is,
\begin{equation} \label{equ:f_F}
 f_{\rm F}=f-f_{\rm G}.
\end{equation}
The $f_{\rm F}$ flux curve of the example data is shown in Figure \ref{fig2}(c). (Note that $f=f_{\rm F}+f_{\rm G}=f_{\rm F}+(f_{\rm M}+f_{\rm L})$; see the illustration in Figure \ref{fig2}.)

It can be seen from Figure \ref{fig2} that the noises in the original \emph{Kepler} light curves are left in the $f_{\rm F}$ flux data. Since the noise level of the \emph{Kepler} data is low \citep{2016RPPh...79c6901B, KeplerDataHandbook} and the flare signal is high (see Figure \ref{fig2}(c)), it will not affect the analysis of flare activities. The specific noise level values of the $f_{\rm F}$ flux data are evaluated in Appendix \ref{sec:appendix_noise_level}.

\subsection{Light Curve Data after Processing}

\begin{figure}
  \epsscale{0.9}
  \plotone{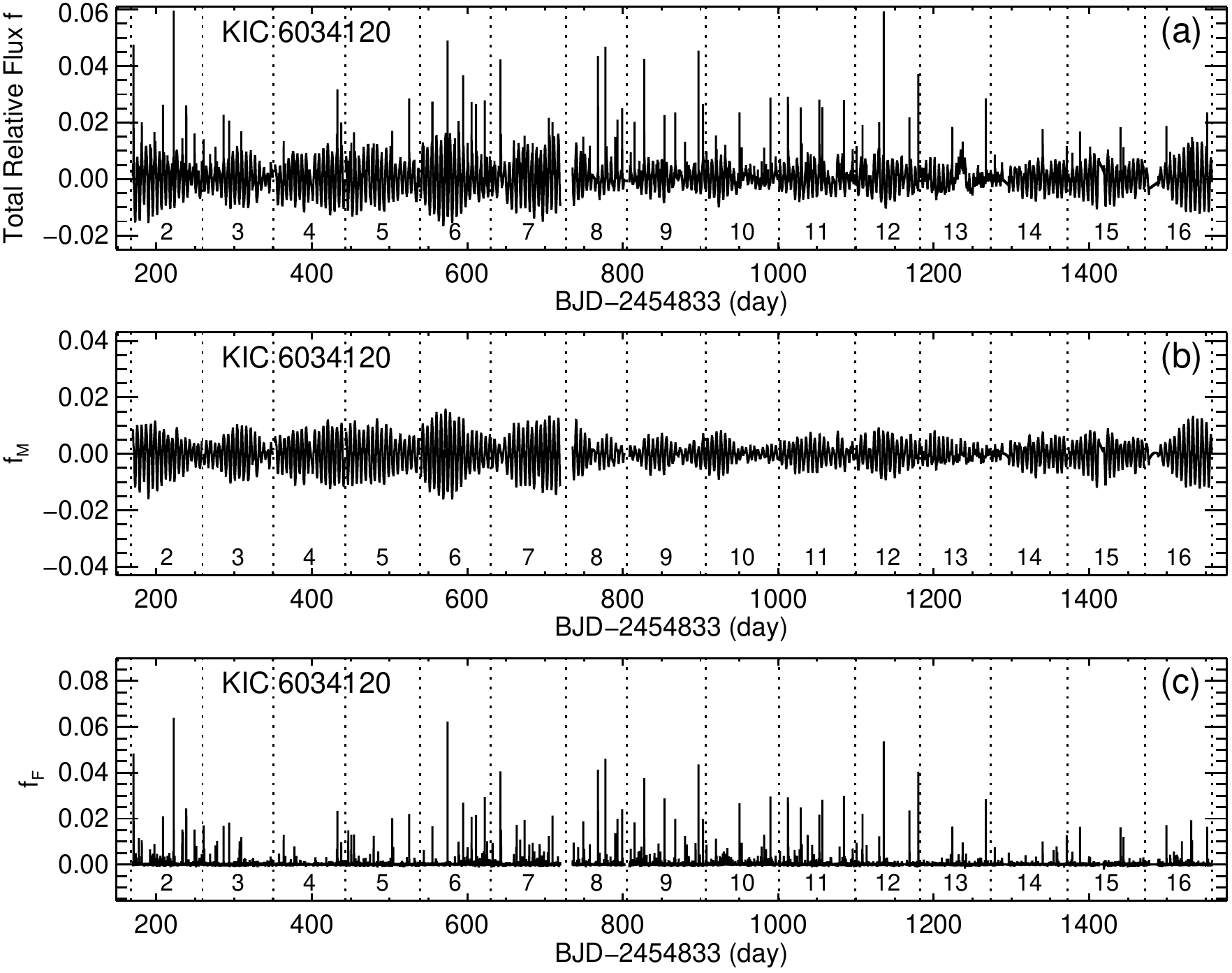}
  \caption{Overview of the processed data (total relative flux $f$, rotational modulation component $f_{\rm M}$, and flare component $f_{\rm F}$) in Q2--Q16 for the star KIC 6034120. \label{fig3}}
\end{figure}
\begin{figure}
  \epsscale{0.9}
  \plotone{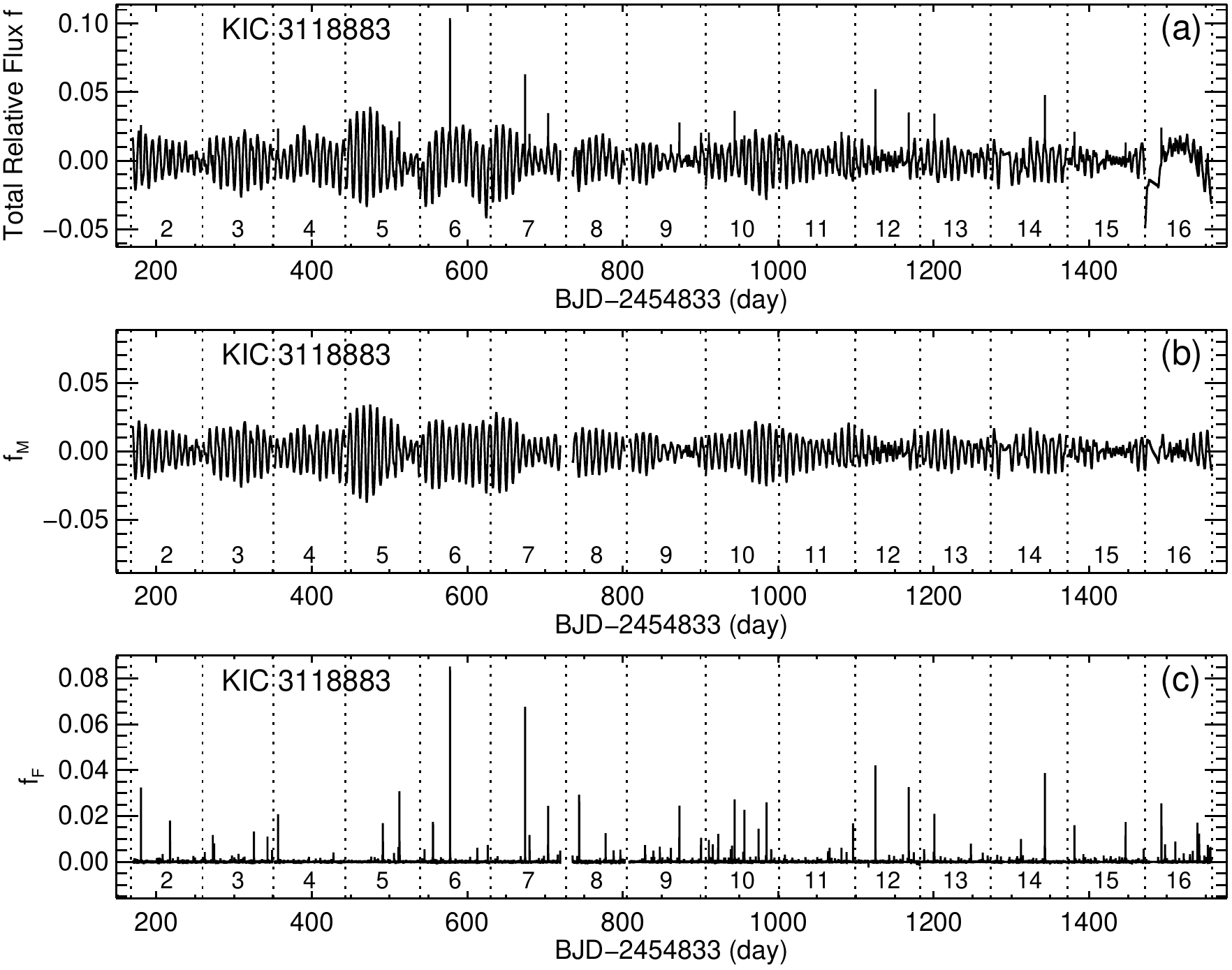}
  \caption{Same as Figure \ref{fig3}, but for the star KIC 3118883. \label{fig4}}
\end{figure}
\begin{figure}
  \epsscale{0.9}
  \plotone{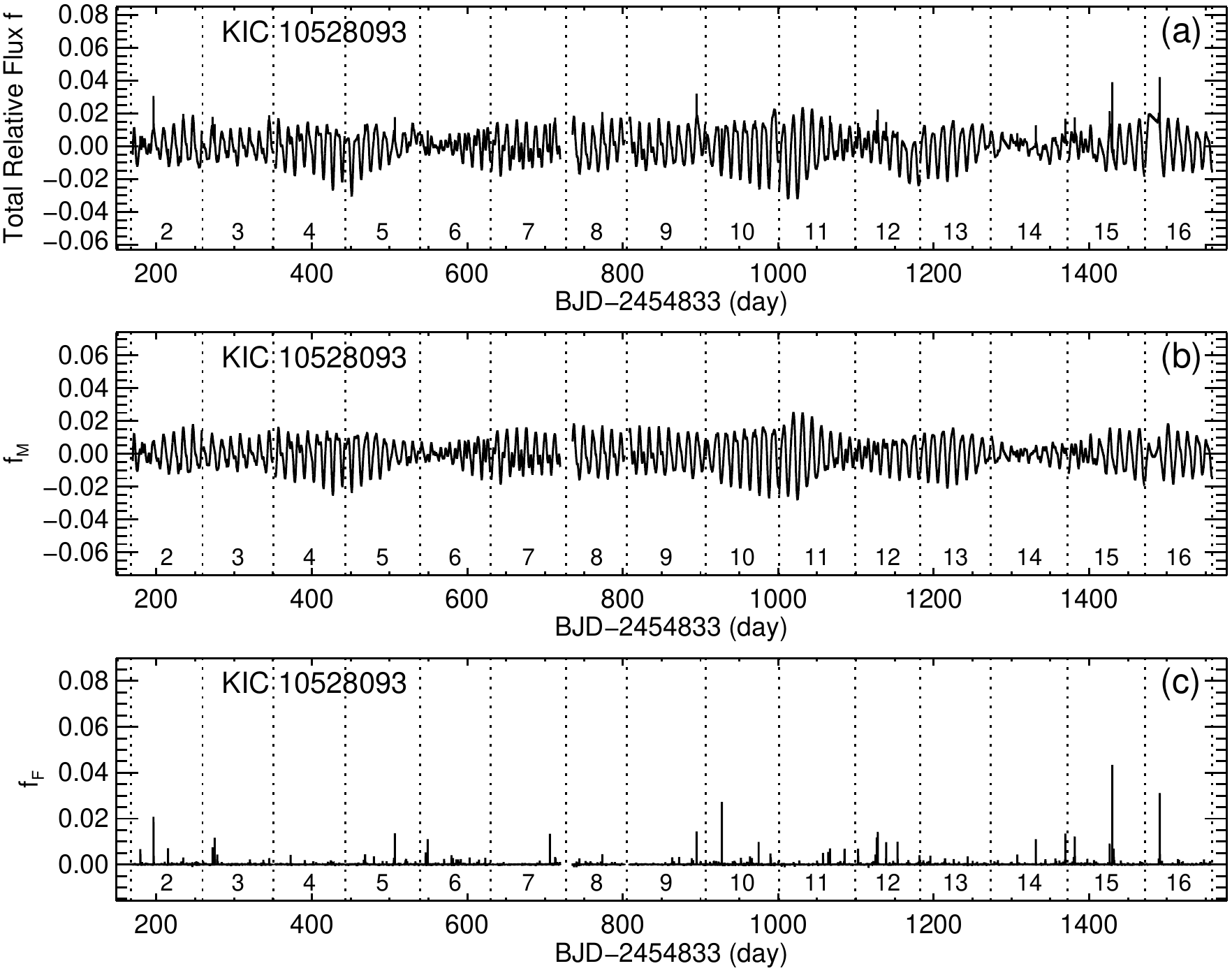}
  \caption{Same as Figure \ref{fig3}, but for the star KIC 10528093. \label{fig5}}
\end{figure}

We processed all the light curve data (in Q2--Q16) of the three investigated stars using the procedure described above. An overview of the processed data is given in Figures \ref{fig3}, \ref{fig4}, and \ref{fig5} for the three stars, KIC 6034120, KIC 3118883, and KIC 10528093, respectively. Figures \ref{fig3}(a), \ref{fig4}(a), and \ref{fig5}(a) show the curves of the total relative flux $f$ of the three stars (i.e., the original \emph{Kepler} light curves in relative flux expression). The missing data points in $f$ within each quarter are patched up through linear interpolation (see Section \ref{subsec:f_M}), and the data gaps between quarters are left over, since we process and analyze the \emph{Kepler} light curve data quarter by quarter. Figures \ref{fig3}(b), \ref{fig4}(b), and \ref{fig5}(b) display the $f_{\rm M}$ (rotational modulation component) flux curves of the three stars and Figures \ref{fig3}(c), \ref{fig4}(c), and \ref{fig5}(c) show the $f_{\rm F}$ (flare component) flux curves. The rotational modulation component $f_{\rm M}$ reflects the activity properties of magnetic features as demonstrated in Paper I, and the flare component $f_{\rm F}$ reflects the activity properties of flares. It can be seen from Figures \ref{fig3}, \ref{fig4}, and \ref{fig5} that the behaviors of both $f_{\rm M}$ and $f_{\rm F}$ components vary with time, which reflects the time evolutions of both magnetic feature and flare activities. To quantitatively describe the activity properties revealed by the $f_{\rm M}$ and $f_{\rm F}$ flux curves, in Section \ref{sec:parameters} we employ magnetic proxies based on $f_{\rm M}$ flux data and flare indexes based on $f_{\rm F}$ flux data, and evaluate the values of the parameters quarter by quarter. Then, the relations of the magnetic feature activities and the flare activities are analyzed using these quantitative parameters in Section \ref{sec:relations}.

\section{Magnetic Proxies and Flare Indexes} \label{sec:parameters}

\subsection{Magnetic Proxies} \label{subsec:proxies}
We use the two magnetic proxies $i_{\rm AC}$ and $R_{\rm eff}$, suggested in Paper I, to quantitatively describe the magnetic activity properties revealed by the $f_{\rm M}$ flux data (rotational modulation component of the total relative flux $f$) derived in Section \ref{sec:data}. The first parameter $i_{\rm AC}$ (autocorrelation index) measures the degree of periodicity of a light curve, which is defined through the autocorrelation algorithm \citep[e.g.,][]{Chatfield-2003} and reflects the stability of the magnetic features that dominate the rotational modulation. The second parameter $R_{\rm eff}$ measures the effective range of light curve fluctuation, which is defined through the rms algorithm \citep[e.g.,][]{2010Sci...329.1032G, 2011ApJ...732L...5C} and reflects the spatial size or coverage of magnetic features \citep[e.g.,][]{2010ApJ...713L.155B, 2011AJ....141...20B, 2013ApJ...769...37B, 2012A&A...539A.137M, 2013MNRAS.432.1203M, 2017A&A...603A..52R}. $R_{\rm eff}$ is a heritage of the light curve range concept commonly used in the astronomy community. The equations for computing the values of $i_{\rm AC}$ and $R_{\rm eff}$ and the approaches for evaluating the errors of the two parameters can be found in Appendix \ref{sec:appendix_error_estimates}.

We calculated the values of $i_{\rm AC}$ and $R_{\rm eff}$, with their errors $\epsilon_i$ and $\epsilon_R$, in each quarter of Q2--Q16, for the three investigated stars from the $f_{\rm M}$ flux data. The results are listed in Table \ref{tab:iac-reff}. The mean values of $i_{\rm AC}$, $R_{\rm eff}$ and $\epsilon_i$, $\epsilon_R$ within the 15 quarters were also calculated and are given in the bottom line of Table \ref{tab:iac-reff} for reference.

\floattable
\begin{deluxetable}{lccccccccccccccc}
\tablecaption{Values of $i_{\rm AC}$ and $R_{\rm eff}$ with Their Errors $\epsilon_i$ and $\epsilon_R$ in Each Quarter of Q2--Q16 for the Three Investigated Stars \label{tab:iac-reff}}
\tablehead{
\colhead{Quarter}  &&  \multicolumn{4}{c}{KIC 6034120}  &&  \multicolumn{4}{c}{KIC 3118883}  &&  \multicolumn{4}{c}{KIC 10528093} \\
\cline{3-6} \cline{8-11} \cline{13-16}
 &&  \colhead{$i_{\rm AC}$} & \colhead{$\epsilon_i$} & \colhead{$R_{\rm eff}$} & \colhead{$\epsilon_R$}
 &&  \colhead{$i_{\rm AC}$} & \colhead{$\epsilon_i$} & \colhead{$R_{\rm eff}$} & \colhead{$\epsilon_R$}
 &&  \colhead{$i_{\rm AC}$} & \colhead{$\epsilon_i$} & \colhead{$R_{\rm eff}$} & \colhead{$\epsilon_R$} \\
 &&                         &                        & $(10^{-2})$             & $(10^{-2})$
 &&                         &                        & $(10^{-2})$             & $(10^{-2})$
 &&                         &                        & $(10^{-2})$             & $(10^{-2})$
}
\startdata
  Q2  &&  0.416 & 0.003 & 1.739 & 0.014  &&  0.374 & 0.005 & 2.821 & 0.034  &&  0.375 & 0.007 & 2.076 & 0.036 \\
  Q3  &&  0.400 & 0.003 & 1.326 & 0.011  &&  0.441 & 0.006 & 3.446 & 0.042  &&  0.371 & 0.007 & 1.765 & 0.030 \\
  Q4  &&  0.413 & 0.003 & 1.768 & 0.014  &&  0.438 & 0.005 & 3.013 & 0.036  &&  0.368 & 0.007 & 2.871 & 0.049 \\
  Q5  &&  0.438 & 0.003 & 1.723 & 0.013  &&  0.407 & 0.005 & 4.858 & 0.055  &&  0.395 & 0.007 & 2.288 & 0.037 \\
  Q6  &&  0.459 & 0.004 & 2.219 & 0.018  &&  0.453 & 0.006 & 4.309 & 0.052  &&  0.220 & 0.004 & 1.414 & 0.024 \\
  Q7  &&  0.373 & 0.003 & 1.978 & 0.016  &&  0.339 & 0.004 & 3.565 & 0.043  &&  0.318 & 0.006 & 2.365 & 0.040 \\
  Q8  &&  0.279 & 0.003 & 1.098 & 0.012  &&  0.454 & 0.008 & 2.801 & 0.045  &&  0.423 & 0.010 & 2.639 & 0.060 \\
  Q9  &&  0.353 & 0.003 & 0.885 & 0.007  &&  0.264 & 0.003 & 1.941 & 0.022  &&  0.387 & 0.006 & 2.411 & 0.038 \\
 Q10  &&  0.259 & 0.002 & 0.882 & 0.007  &&  0.426 & 0.005 & 3.304 & 0.038  &&  0.430 & 0.007 & 3.396 & 0.056 \\
 Q11  &&  0.346 & 0.003 & 0.986 & 0.007  &&  0.338 & 0.004 & 2.561 & 0.029  &&  0.417 & 0.007 & 3.551 & 0.056 \\
 Q12  &&  0.436 & 0.004 & 1.233 & 0.011  &&  0.211 & 0.003 & 1.690 & 0.022  &&  0.349 & 0.007 & 2.111 & 0.039 \\
 Q13  &&  0.290 & 0.002 & 0.723 & 0.006  &&  0.429 & 0.005 & 2.242 & 0.027  &&  0.452 & 0.008 & 2.669 & 0.045 \\
 Q14  &&  0.339 & 0.003 & 0.958 & 0.007  &&  0.368 & 0.004 & 2.410 & 0.027  &&  0.240 & 0.004 & 1.012 & 0.016 \\
 Q15  &&  0.360 & 0.003 & 1.255 & 0.009  &&  0.197 & 0.002 & 1.400 & 0.016  &&  0.351 & 0.006 & 2.228 & 0.035 \\
 Q16  &&  0.403 & 0.003 & 1.666 & 0.014  &&  0.207 & 0.003 & 1.744 & 0.022  &&  0.422 & 0.008 & 2.406 & 0.043 \\
 \cline{1-16}
      &&  \multicolumn{14}{c}{Mean Value} \\
          \cline{3-16}
      &&  0.371 & 0.003 & 1.363 & 0.011  &&  0.356 & 0.005 & 2.807 & 0.034  &&  0.368 & 0.007 & 2.347 & 0.040 \\
\enddata
\end{deluxetable}

\subsection{Flare Indexes} \label{subsec:indexes}
The flare signals in Figures \ref{fig3}(c), \ref{fig4}(c), and \ref{fig5}(c) show variations of flare occurrence frequency, as well as flare magnitude for the three investigated stars. To quantitatively describe flare activity properties, we introduce three flare indexes based on the $f_{\rm F}$ flux data. The three flare indexes are time occupation ratio of flares $R_{\rm flare}$, total relative power (energy rate) of flares $P_{\rm flare}$, and averaged relative flux magnitude of flares $M_{\rm flare}$. These flare indexes are inspired by the analogous measures of solar flares \citep[e.g.,][]{2012ApJ...754..112A, 2015LRSP...12....4H}, and reflect different aspects (frequency, energy, and magnitude) of flare activity properties. Because the three flare indexes are defined as ratio or relative values (see definitions below), they are all dimensionless quantities. As done for the two magnetic proxies in Section \ref{subsec:proxies}, we intend to evaluate the three flare indexes quarter by quarter to exhibit the time evolution information of flare activities.

\begin{figure}
  \epsscale{0.86}
  \plotone{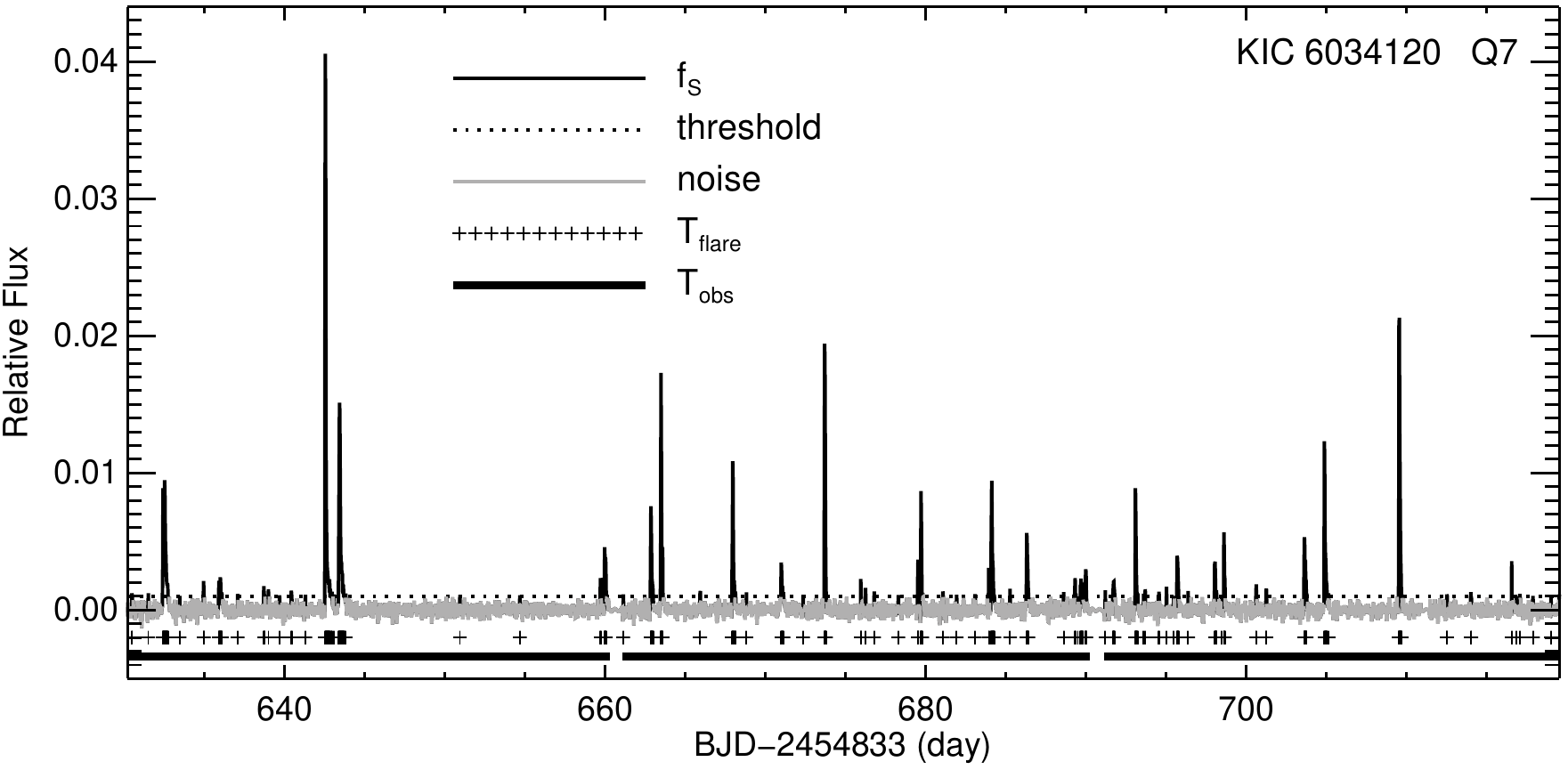}
  \caption{Diagram illustrating the quantities employed in the flare index definitions. The whole flux curve is $f_{\rm F}$, as shown in Figure \ref{fig2}(c). The vertical black spikes are the identified flare signals $f_{\rm S}$. The threshold ($=0.001$) for identifying the flare spikes is indicated by a horizontal dotted line. The gray curve shows the background noises. The occupation time of the flare spikes ($T_{\rm flare}$) is indicated by the plus symbols. The valid observing time of the quarter ($T_{\rm obs}$) is indicated by segments of bold lines. (The gaps between the time segments of $T_{\rm obs}$ are for spacecraft data downlinks.) The example data are from Q7 of KIC 6034120. \label{fig6}}
\end{figure}

In Figure \ref{fig6}, we use the $f_{\rm F}$ flux curve of KIC 6034120 in Q7 (as shown in Figure \ref{fig2}(c)) to explain the quantities employed in the flare index definitions. Firstly, the subset of flare spikes (denoted by $f_{\rm S}$, where the letter `S' refers to `spikes') in $f_{\rm F}$ flux data are identified by the criterion $f_{\rm F} \geqslant 0.001$. The threshold value 0.001 is determined empirically and is just above the noise floor in the light curve data of KIC 6034120 (see the diagram illustration in Figure \ref{fig6}, in which the background noises are shown in gray, and the flare threshold is indicated by a horizontal dotted line). The quantitative noise level values ($\sigma_{\rm N}$) of the light curve data are evaluated in Appendix \ref{sec:appendix_noise_level}. For KIC 6034120, $\sigma_{\rm N}=0.00035$ (see the evaluation details in Appendix \ref{sec:appendix_noise_level}), and the threshold value 0.001 employed to identify the flare spikes roughly equals to $3\times \sigma_{\rm N}$. The other two stars investigated in this paper are brighter than KIC 6034120 (see $Kp$ values in Table \ref{tab:star-param}), and thus have lower noise levels (see the quantitative results in Appendix \ref{sec:appendix_noise_level}). Since we want to compare flare activity properties between stars, we adopt the same flare threshold (i.e., $f_{\rm F} \geqslant 0.001$) for all three investigated stars in this work.

After identification of the flare spikes, the total occupation time of the flares (denoted by $T_{\rm flare}$) can be counted accordingly (see plus symbols in Figure \ref{fig6}). The total valid observing time of the quarter (as recorded in the \emph{Kepler} time-series data) is denoted by $T_{\rm obs}$ (illustrated by segments of bold lines in Figure \ref{fig6}). Note that the time gaps between the three months of the quarter \citep[for spacecraft data downlink;][]{2016RPPh...79c6901B, KeplerDataHandbook} are not included in $T_{\rm obs}$.

The time occupation ratio of flares, $R_{\rm flare}$, is defined as the ratio of $T_{\rm flare}$ to $T_{\rm obs}$,
\begin{equation} \label{equ:Rflare}
  R_{\rm flare}=T_{\rm flare}/T_{\rm obs}.
\end{equation}
Note that the concept of $R_{\rm flare}$ is equivalent to the concept of flare occurrence frequency. We adopt $R_{\rm flare}$ in this work because it is more convenient for evaluation.

For defining the total relative power (energy rate) of flares, $P_{\rm flare}$, we first evaluate the total relative energy (denoted by $U_{\rm flare}$) released by all the identified flares in the quarter via equation
\begin{equation} \label{equ:Uflare}
  U_{\rm flare}=\int f_{\rm S}(t) dt.
\end{equation}
The integral in equation (\ref{equ:Uflare}) is over the discrete time intervals occupied by the identified flare spikes, which are indicated by the plus symbols in Figure 6. Then, the total relative power (energy rate) of flares, $P_{\rm flare}$, can be calculated by equation
\begin{equation} \label{equ:Pflare}
  P_{\rm flare}=U_{\rm flare}/T_{\rm obs}.
\end{equation}
Note that $U_{\rm flare}$ has the dimension of time \citep[and is also called the equivalent duration of flare in the literature, e.g.,][]{2014ApJ...797..121H} and $P_{\rm flare}$ is dimensionless. Since $f_{\rm F}$ (and hence $f_{\rm S}$) is evaluated relative to the background flux of the star (see section \ref{sec:data}), the value of $P_{\rm flare}$ is also relative to the background energy emission rate of the star.

The averaged relative flux magnitude of flares, $M_{\rm flare}$, is defined by equation
\begin{equation} \label{equ:Mflare}
  M_{\rm flare}=\frac{1}{T_{\rm flare}} \int f_{\rm S}(t) dt=U_{\rm flare}/T_{\rm flare}.
\end{equation}

It can be seen from equations (\ref{equ:Rflare}), (\ref{equ:Pflare}), and (\ref{equ:Mflare}) that the three flare indexes $M_{\rm flare}$, $R_{\rm flare}$, and $P_{\rm flare}$ are connected by equation
\begin{equation} \label{equ:RPMflare}
  M_{\rm flare}=P_{\rm flare}/R_{\rm flare}.
\end{equation}
The approaches for evaluating the errors of the three flare indexes can be found in Appendix \ref{sec:appendix_error_estimates_flare}.

\floattable
\begin{deluxetable}{lccccccccccccccccccccc}
\tablecaption{Values of $R_{\rm flare}$, $P_{\rm flare}$, and $M_{\rm flare}$, with Their Errors $\epsilon^{R}_{\rm flare}$, $\epsilon^{P}_{\rm flare}$, and $\epsilon^{M}_{\rm flare}$, in Each Quarter of Q2--Q16 for the Three Investigated Stars \label{tab:flare-indexes}}
\rotate
\tablehead{
\colhead{Quarter}  &&  \multicolumn{6}{c}{KIC 6034120}  &&  \multicolumn{6}{c}{KIC 3118883}  &&  \multicolumn{6}{c}{KIC 10528093} \\
\cline{3-8} \cline{10-15} \cline{17-22}
 &&  \colhead{$R_{\rm flare}$}  &  \colhead{$\epsilon^{R}_{\rm flare}$}  &  \colhead{$P_{\rm flare}$}  &  \colhead{$\epsilon^{P}_{\rm flare}$}  &
     \colhead{$M_{\rm flare}$}  &  \colhead{$\epsilon^{M}_{\rm flare}$}
 &&  \colhead{$R_{\rm flare}$}  &  \colhead{$\epsilon^{R}_{\rm flare}$}  &  \colhead{$P_{\rm flare}$}  &  \colhead{$\epsilon^{P}_{\rm flare}$}  &
     \colhead{$M_{\rm flare}$}  &  \colhead{$\epsilon^{M}_{\rm flare}$}
 &&  \colhead{$R_{\rm flare}$}  &  \colhead{$\epsilon^{R}_{\rm flare}$}  &  \colhead{$P_{\rm flare}$}  &  \colhead{$\epsilon^{P}_{\rm flare}$}  &
     \colhead{$M_{\rm flare}$}  &  \colhead{$\epsilon^{M}_{\rm flare}$}  \\
 &&  $(10^{-2})$  &  $(10^{-2})$  &  $(10^{-4})$  & $(10^{-4})$  &  $(10^{-3})$  &  $(10^{-3})$
 &&  $(10^{-2})$  &  $(10^{-2})$  &  $(10^{-4})$  & $(10^{-4})$  &  $(10^{-3})$  &  $(10^{-3})$
 &&  $(10^{-2})$  &  $(10^{-2})$  &  $(10^{-4})$  & $(10^{-4})$  &  $(10^{-3})$  &  $(10^{-3})$
}
\startdata
 Q2 && 6.295 & 0.024 & 2.721 & 0.014 & 4.322 & 0.022 && 1.502 & 0.024 & 0.576 & 0.005 & 3.834 & 0.034 && 1.335 & 0.024 & 0.523 & 0.003 & 3.915 & 0.021\\
 Q3 && 4.210 & 0.024 & 1.292 & 0.011 & 3.069 & 0.026 && 1.679 & 0.024 & 0.465 & 0.005 & 2.769 & 0.032 && 1.159 & 0.024 & 0.369 & 0.003 & 3.184 & 0.023\\
 Q4 && 2.525 & 0.024 & 0.743 & 0.009 & 2.942 & 0.034 && 0.745 & 0.024 & 0.252 & 0.004 & 3.384 & 0.048 && 0.385 & 0.024 & 0.061 & 0.002 & 1.597 & 0.040\\
 Q5 && 3.129 & 0.022 & 1.097 & 0.009 & 3.505 & 0.029 && 1.476 & 0.022 & 0.704 & 0.005 & 4.767 & 0.033 && 1.388 & 0.022 & 0.317 & 0.003 & 2.284 & 0.020\\
 Q6 && 5.379 & 0.023 & 2.152 & 0.012 & 4.001 & 0.023 && 2.017 & 0.023 & 1.192 & 0.006 & 5.910 & 0.029 && 1.182 & 0.023 & 0.305 & 0.003 & 2.583 & 0.022\\
 Q7 && 6.310 & 0.023 & 2.332 & 0.013 & 3.697 & 0.021 && 1.746 & 0.023 & 1.081 & 0.005 & 6.192 & 0.031 && 0.233 & 0.023 & 0.064 & 0.001 & 2.749 & 0.051\\
 Q8 && 6.077 & 0.032 & 2.732 & 0.015 & 4.496 & 0.025 && 2.450 & 0.032 & 0.811 & 0.008 & 3.310 & 0.031 && 0.573 & 0.032 & 0.103 & 0.002 & 1.799 & 0.038\\
 Q9 && 6.379 & 0.021 & 2.648 & 0.013 & 4.151 & 0.020 && 1.573 & 0.021 & 0.525 & 0.005 & 3.338 & 0.031 && 0.936 & 0.021 & 0.209 & 0.002 & 2.237 & 0.024\\
Q10 && 6.992 & 0.022 & 1.871 & 0.014 & 2.676 & 0.020 && 2.961 & 0.022 & 1.320 & 0.007 & 4.457 & 0.023 && 1.202 & 0.022 & 0.383 & 0.003 & 3.182 & 0.022\\
Q11 && 6.031 & 0.022 & 2.161 & 0.013 & 3.584 & 0.021 && 1.732 & 0.022 & 0.518 & 0.005 & 2.991 & 0.030 && 1.075 & 0.022 & 0.300 & 0.002 & 2.793 & 0.023\\
Q12 && 3.531 & 0.025 & 1.806 & 0.010 & 5.114 & 0.030 && 1.463 & 0.025 & 0.902 & 0.005 & 6.164 & 0.035 && 2.043 & 0.025 & 0.774 & 0.004 & 3.790 & 0.018\\
Q13 && 2.395 & 0.023 & 0.649 & 0.008 & 2.709 & 0.034 && 1.865 & 0.023 & 0.599 & 0.006 & 3.211 & 0.030 && 0.484 & 0.023 & 0.079 & 0.002 & 1.637 & 0.035\\
Q14 && 1.963 & 0.023 & 0.557 & 0.007 & 2.840 & 0.038 && 1.737 & 0.023 & 0.629 & 0.005 & 3.619 & 0.031 && 1.421 & 0.023 & 0.400 & 0.003 & 2.814 & 0.020\\
Q15 && 2.380 & 0.022 & 0.634 & 0.008 & 2.664 & 0.034 && 1.935 & 0.022 & 0.525 & 0.006 & 2.715 & 0.029 && 3.159 & 0.022 & 1.210 & 0.004 & 3.831 & 0.013\\
Q16 && 3.706 & 0.028 & 1.209 & 0.011 & 3.263 & 0.030 && 3.176 & 0.028 & 1.215 & 0.008 & 3.824 & 0.025 && 1.031 & 0.028 & 0.485 & 0.003 & 4.709 & 0.026\\
     \cline{1-22}
     && \multicolumn{20}{c}{Mean Value} \\
     \cline{3-22}
    && 4.487 & 0.024 & 1.640 & 0.011 & 3.536 & 0.027 && 1.870 & 0.024 & 0.754 & 0.006 & 4.032 & 0.031 && 1.174 & 0.024 & 0.372 & 0.003 & 2.874 & 0.026\\
\enddata
\end{deluxetable}

We calculated the values of the three flare indexes, $R_{\rm flare}$, $P_{\rm flare}$, and $M_{\rm flare}$, with their errors $\epsilon^{R}_{\rm flare}$, $\epsilon^{P}_{\rm flare}$, and $\epsilon^{M}_{\rm flare}$, in each quarter of Q2--Q16 for the three investigated stars. The results are listed in Table \ref{tab:flare-indexes}. The mean values of the flare indexes and errors within the 15 quarters are given in the bottom line of Table \ref{tab:flare-indexes} for reference.

\section{Magnetic Feature vs. Flare Activity} \label{sec:relations}

\subsection{Magnetic Feature Activity} \label{subsec:mag-activity}

\begin{figure}
  \epsscale{0.88}
  \plotone{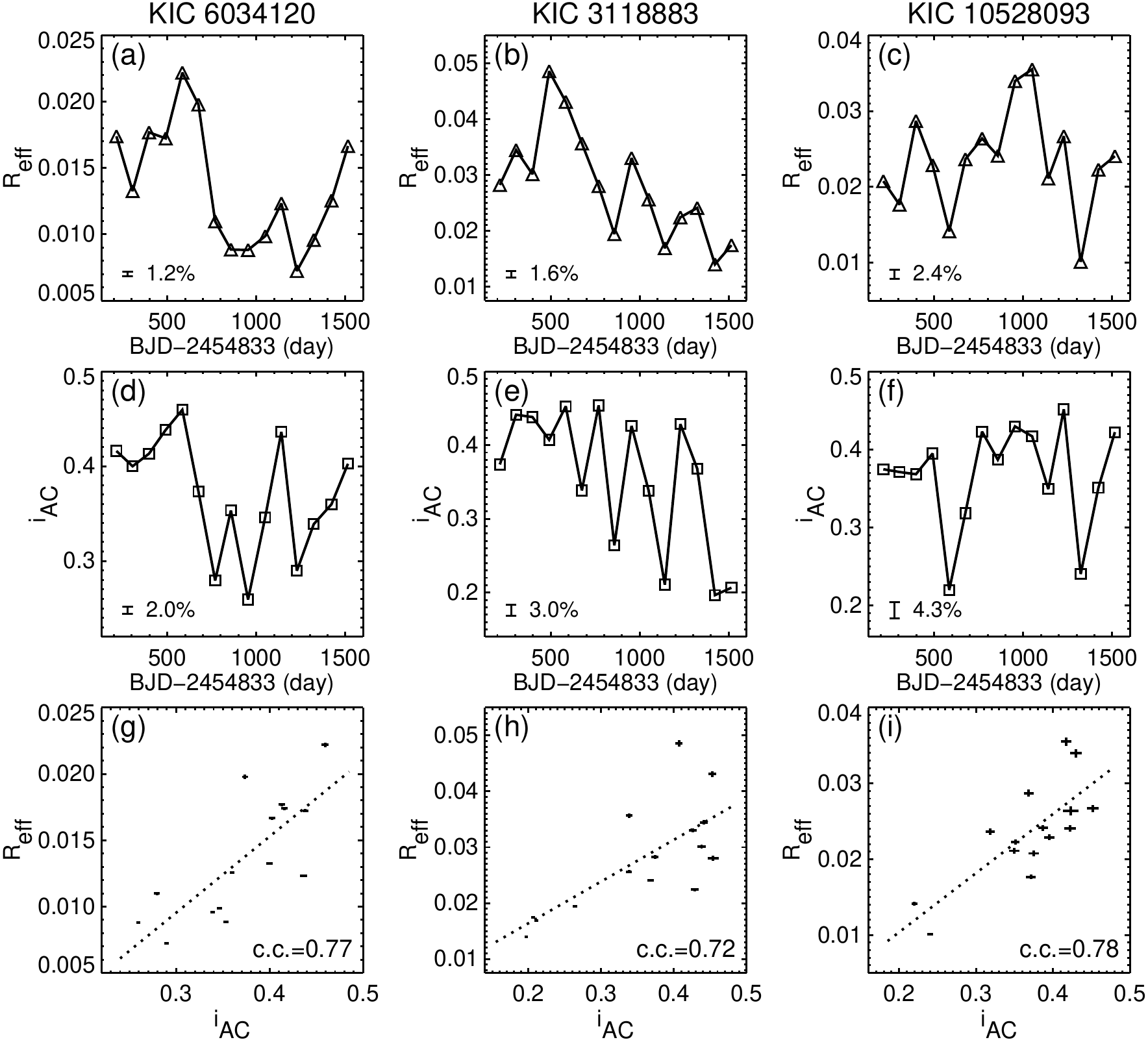}
  \caption{Variations of $R_{\rm eff}$ and $i_{\rm AC}$ with time for the three investigated stars (top and middle rows) and correlation analysis (scatter plot, correlation coefficient, and linear fitting) for the paired data of the two magnetic proxies (bottom row). Each column corresponds to one star, and each data point in the plots corresponds to one quarter in Q2--Q16. In the top and middle rows, the time coordinate of each data point is the central time of the corresponding quarter; the error bars in the lower left corner of each panel depict the largest value of $\pm\epsilon_R$ or $\pm\epsilon_i$ (see $\epsilon_R$ and $\epsilon_i$ columns in Table \ref{tab:iac-reff}) among the 15 quarters of each plot; and the percentage value by the error bar is the ratio of the error bar height to the variation range of $R_{\rm eff}$ or $i_{\rm AC}$ (see the main text). In the bottom row, the dotted lines are linear fitting for the paired data of $i_{\rm AC}$ and $R_{\rm eff}$; `c.c.' refers to correlation coefficient (see equation (\ref{equ:corr})); and the vertical and horizontal lengths of the plus symbols are drawn according to the $\pm\epsilon_R$ and $\pm\epsilon_i$ values of the paired data (see Table \ref{tab:iac-reff}). \label{fig7}}
\end{figure}

As explained in section \ref{subsec:proxies}, the two magnetic proxies $i_{\rm AC}$ and $R_{\rm eff}$ quantitatively characterize the rotation modulation component $f_{\rm M}$ of \emph{Kepler} light curves caused by the corotating magnetic features on stellar surface, thus the two parameters can reflect the magnetic feature activity properties of stars. The variations of the two magnetic proxies with time for the three investigated stars KIC 6034120, KIC 3118883, and KIC 10528093 are plotted in Figure \ref{fig7} (top row for $R_{\rm eff}$, middle row for $i_{\rm AC}$, and each column for one star). The values of the two quantities are taken from Table \ref{tab:iac-reff}. Each data point in the plots corresponds to one quarter in Q2--Q16. (We denote the values of $R_{\rm eff}$ and $i_{\rm AC}$ of different quarters by $R_{\rm eff}^q$ and $i_{\rm AC}^q$, where $q=2, 3, \ldots, 16$ is the quarter number. The time coordinate of each data point is the central time of the corresponding quarter.) Because the errors of $R_{\rm eff}$ and $i_{\rm AC}$ for the three investigated stars are relatively small (see $\epsilon_R$ and $\epsilon_i$ columns in Table \ref{tab:iac-reff}), in Figure \ref{fig7} (top and middle rows), we use one error bar in the lower left corner of each panel to indicate the largest value of $\epsilon_R$ or $\epsilon_i$ of each plot (note that a error bar depicts $\pm\epsilon_R$ or $\pm\epsilon_i$ and its height equals to $2\epsilon_R$ or $2\epsilon_i$). The percentage value of the error bar height to the variation range of $R_{\rm eff}$ or $i_{\rm AC}$ (defined by $\max[\{R_{\rm eff}^q\}]-\min[\{R_{\rm eff}^q\}]$ or $\max[\{i_{\rm AC}^q\}]-\min[\{i_{\rm AC}^q\}]$) is shown by the error bar in each panel.

It can be seen from the plots in the top and middle rows of Figure \ref{fig7} that both the $R_{\rm eff}$ and $i_{\rm AC}$ show long length-scale variations, and the variation ranges of the two magnetic proxies are much larger than their errors, as indicated by the error bars and the percentage values in the lower left corner of each plot. Additionally, the long length-scale variations of the two magnetic proxies are in the same phase for all three stars. To verify this impression, we performed a correlation analysis for the paired data of $i_{\rm AC}$ and $R_{\rm eff}$. The scatter plots of the ($i_{\rm AC}^q$, $R_{\rm eff}^q$) pairs, the correlation coefficients, and the linear fitting for the paired data of the three stars are given in the bottom row of Figure \ref{fig7}. Here, the correlation coefficient (abbreviated as `c.c.') between the paired data of $i_{\rm AC}$ and $R_{\rm eff}$ of each star is calculated using the following formula:
\begin{equation}\label{equ:corr}
  {\rm c.c.}=\frac
             {\sum_{q=2}^{16}\left({i_{\rm AC}^{q}-\overline{i_{\rm AC}}}\right)\left({R_{\rm eff}^{q}-\overline{R_{\rm eff}}}\right)}
             {\sqrt{\sum_{q=2}^{16}\left({i_{\rm AC}^{q}-\overline{i_{\rm AC}}}\right)^2}
              \sqrt{\sum_{q=2}^{16}\left({R_{\rm eff}^{q}-\overline{R_{\rm eff}}}\right)^2}},
\end{equation}
where $\overline{i_{\rm AC}}$ and $\overline{R_{\rm eff}}$ are mean values of $\{i_{\rm AC}^q\}$ and $\{R_{\rm eff}^q\}$. The results of the correlation analysis affirm that the two magnetic proxies are positively correlated for the three stars, with the correlation coefficients for the three stars being 0.77, 0.72, and 0.78, respectively. So all three stars being investigated in this paper belong to the positive-correlation-star category described in the work by Mehrabi et al. (\citeyear{2017ApJ...834..207M}; see the introduction in Section \ref{sec:intro}).

\subsection{Flare Activity} \label{subsec:flare-activity}

\begin{figure}
  \epsscale{0.88}
  \plotone{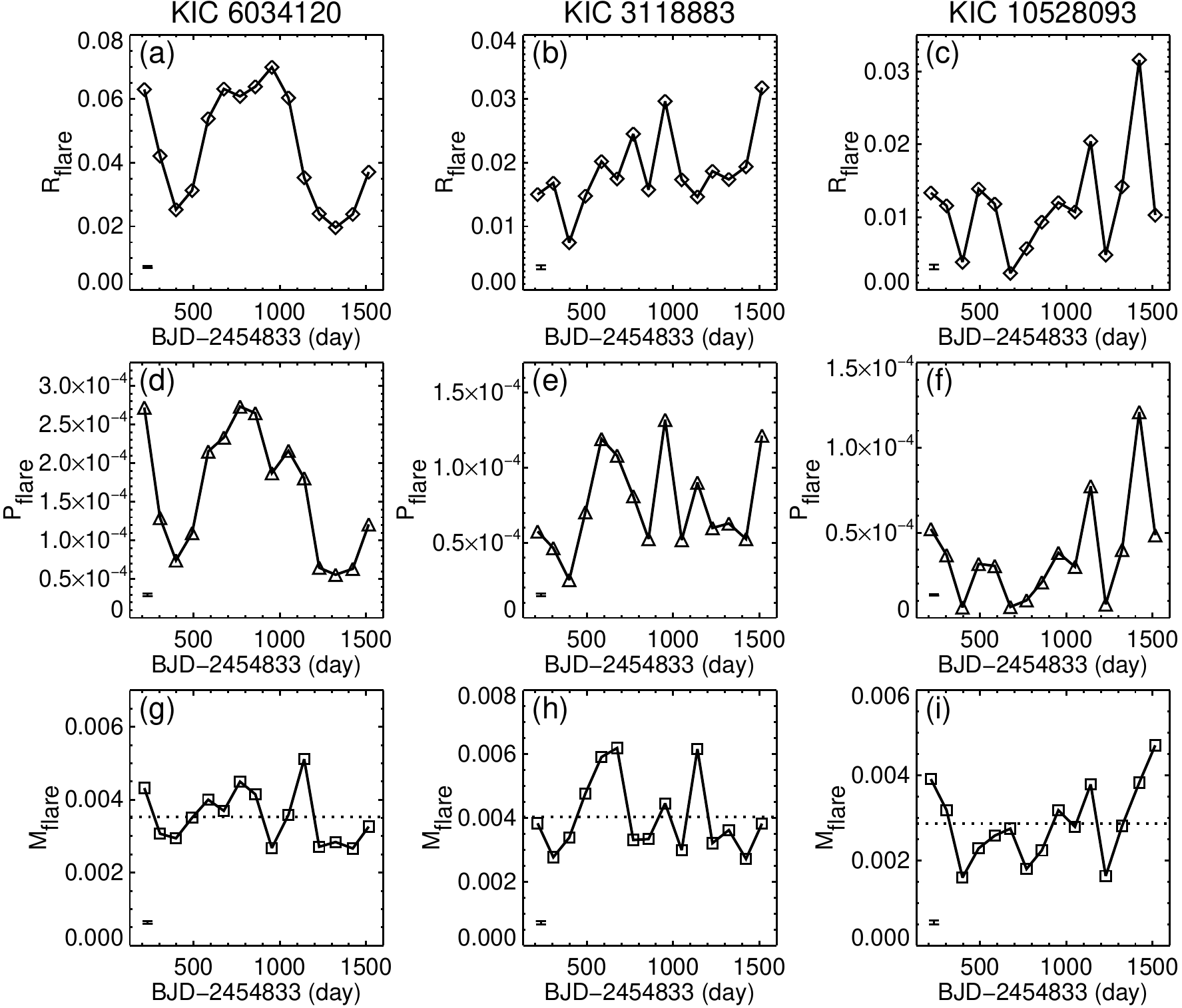}
  \caption{Variations of $R_{\rm flare}$ (top row), $P_{\rm flare}$ (middle row), and $M_{\rm flare}$ (bottom row) with time for the three investigated stars. Each column corresponds to one star, and each data point in the plots corresponds to one quarter in Q2--Q16. The time coordinate of each data point is the central time of the corresponding quarter. The dotted lines in the bottom row indicate the mean values of $M_{\rm flare}$. The error bars in the lower left corner of each panel depict the largest value of $\pm\epsilon^{R}_{\rm flare}$, $\pm\epsilon^{P}_{\rm flare}$, or $\pm\epsilon^{M}_{\rm flare}$ (see $\epsilon^{R}_{\rm flare}$, $\epsilon^{P}_{\rm flare}$, and $\epsilon^{M}_{\rm flare}$ columns in Table \ref{tab:flare-indexes}) among the 15 quarters of each plot.  \label{fig8}}
\end{figure}

Variations of the three flare indexes with time for the three investigated stars are plotted in Figure \ref{fig8} (top row for $R_{\rm flare}$, middle row for $P_{\rm flare}$, bottom row for $M_{\rm flare}$, and each column corresponding to one star). The values of the three flare indexes are taken from Table \ref{tab:flare-indexes}. The error bars in the lower left corner of each panel depict the largest value of $\pm\epsilon^{R}_{\rm flare}$, $\pm\epsilon^{P}_{\rm flare}$, or $\pm\epsilon^{M}_{\rm flare}$ (see relevant columns in Table \ref{tab:flare-indexes}) of each plot. (Note that the errors of the three flare indexes are fairly small relative to the variation ranges of the flare indexes.)

It can be seen from Figure \ref{fig8}(a)--(c) that the curves of the flare index $R_{\rm flare}$ (time occupation ratio of flares) show long-term trends of flare activity for the three investigated stars, while the flare index $M_{\rm flare}$ (averaged relative flux magnitude of flares) does not show a long-term trend and fluctuates randomly around its mean value for the three stars (see Figures \ref{fig8}(g)--(i)). (Note that this property of $M_{\rm flare}$ is analogous to the statistical result on the invariant solar flare magnitude distribution throughout the solar cycle \citep{2012ApJ...754..112A, 2015LRSP...12....4H} described in Section \ref{sec:intro}.) The flare index $P_{\rm flare}$ (total relative power of flares) displays similar long-term trends as the flare index $R_{\rm flare}$ (see Figures \ref{fig8}(d)--(f)). Considering $P_{\rm flare}$ is the product of $R_{\rm flare}$ and $M_{\rm flare}$ (see equation (\ref{equ:RPMflare})), and $M_{\rm flare}$ varies around its mean value, the long-term variation of $P_{\rm flare}$ is in fact the reflection of the long-term variation of $R_{\rm flare}$.

Since the mean $R_{\rm flare}$ value of KIC 6034120 is higher than that of KIC 3118883 and KIC 10528093 (see the bottom line in Table \ref{tab:flare-indexes}), i.e., there are more flares on KIC 6034120 than KIC 3118883 and KIC 10528093 (see Figures \ref{fig3}--\ref{fig5}), the statistical significance of the $R_{\rm flare}$ variation for KIC 6034120 is also higher, thus the $R_{\rm flare}$ curve of KIC 6034120 is smoother (less random fluctuation) than the other two stars as presented in Figures \ref{fig8}(a)--(c).

\subsection{Correlation Analysis between Magnetic Feature and Flare Activities} \label{subsec:corr}

\begin{figure}
  \epsscale{0.88}
  \plotone{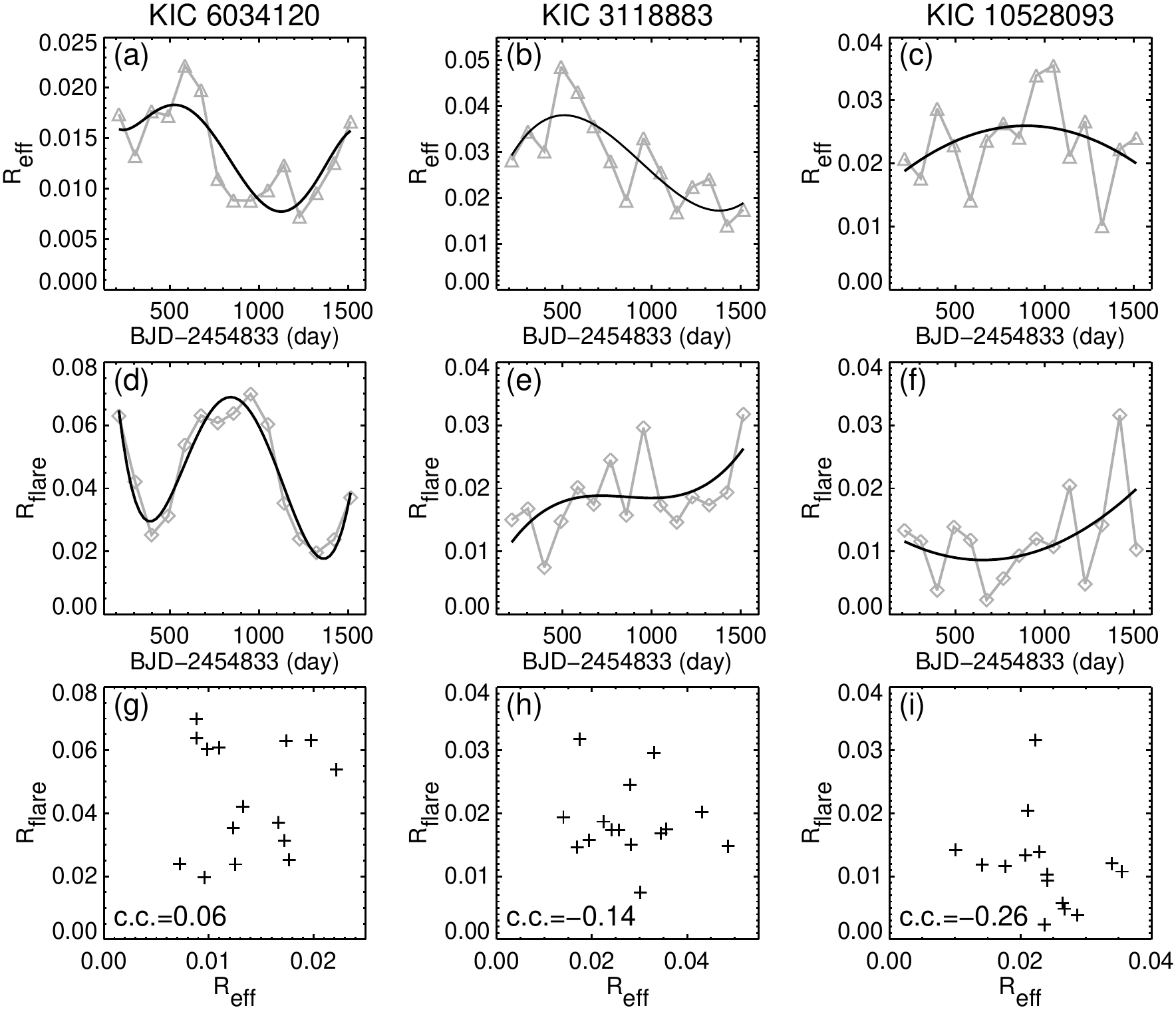}
  \caption{Polynomial fitting for the $R_{\rm eff}$ curves (top row) and $R_{\rm flare}$ curves (middle row) of the three investigated stars and correlation analysis (scatter plot and correlation coefficient) for the paired data of $R_{\rm eff}$ and $R_{\rm flare}$ (bottom row). Each column corresponds to one star, and each data point in the plots corresponds to one quarter in Q2--Q16. In the top and middle rows, the polynomial fitting curves are shown in black and the original $R_{\rm eff}$ and $R_{\rm flare}$ curves are shown in gray. In the bottom row, `c.c.' refers to correlation coefficient. \label{fig9}}
\end{figure}

Because the long length-scale variations of the two magnetic proxies $i_{\rm AC}$ and $R_{\rm eff}$ are in the same phase for the three investigated stars (see Section \ref{subsec:mag-activity}) and the errors of $R_{\rm eff}$ are smaller than the errors of $i_{\rm AC}$ (relative to the variation ranges of the two magnetic proxies; see the percentage values given in the top and middle rows of Figure \ref{fig7}), we adopt $R_{\rm eff}$ (effective range of light curve fluctuation) as the representative parameter of magnetic feature activity for the three investigated stars. The long-term trends of the flare activities of the three stars are best represented by the flare index $R_{\rm flare}$ (time occupation ratio of flares), as demonstrated in Section \ref{subsec:flare-activity}. Then, the relation between the magnetic feature activities and the flare activities of the three stars can be quantitatively described by the correlation between the values of $R_{\rm eff}$ and $R_{\rm flare}$ of the three stars. The results of the correlation analysis between $R_{\rm eff}$ and $R_{\rm flare}$ for the three stars are given in Figure \ref{fig9}.

To indicate the long-term trends of magnetic feature and flare activities more clearly, we performed polynomial fitting for the $R_{\rm eff}$ and $R_{\rm flare}$ curves of the three investigated stars. The results of the polynomial fitting are shown in the top row (for $R_{\rm eff}$) and middle row (for $R_{\rm flare}$) of Figure \ref{fig9} (each column corresponding to one star). The polynomial fitting curves are shown in black and the original $R_{\rm eff}$ and $R_{\rm flare}$ curves are shown in gray. The orders of polynomial fitting were determined empirically, which are 5, 3, and 2 for KIC 6034120, KIC 3118883, and KIC 10528093, respectively. The scatter plots of the paired data of $R_{\rm eff}$ and $R_{\rm flare}$ are given in the bottom row of Figure \ref{fig9}. The correlation coefficients between $R_{\rm eff}$ and $R_{\rm flare}$ of the three stars are $0.06$, $-0.14$, and $-0.26$, respectively. (The formula for calculating the correlation coefficients is analogous to equation (\ref{equ:corr}), except that $i_{\rm AC}$ is replaced with $R_{\rm flare}$.)

It can be seen in the top and middle rows of Figure \ref{fig9} that the long-term variations of $R_{\rm eff}$ and $R_{\rm flare}$ are not in phase with each other for all three stars. This impression is affirmed by the near zero or negative correlations between $R_{\rm eff}$ and $R_{\rm flare}$ as demonstrated in the bottom row of Figure \ref{fig9}. The fact that the magnetic feature activity and the flare activity are not positively related indicates that they may have different source regions.

\subsection{Cycle Length of Magnetic Feature and Flare Activities} \label{subsec:c-length}

The polynomial fitting plots for the star KIC 6034120 in Figures \ref{fig9}(a) and (d) show apparent cyclic variations of both the parameters $R_{\rm eff}$ and $R_{\rm flare}$, which might represent the magnetic feature activity cycle and the flare activity cycle, respectively. Although the cyclic variations of two parameters are not in phase with each other, as demonstrated in Section \ref{subsec:corr}, the rhythms of the variations are compatible (see Figures \ref{fig9}(a) and (d)), that is, they may have similar cycle lengths. The fitting curves for the other two stars KIC 3118883 and KIC 10528093 also show clues of this impression (see Figures \ref{fig9}(b), (c), (e), and (f)), but are not as apparent as KIC 6034120 since the $R_{\rm eff}$ and $R_{\rm flare}$ curves of these two stars do not exhibit a full cycle of variation.

\begin{figure}
  \epsscale{0.58}
  \plotone{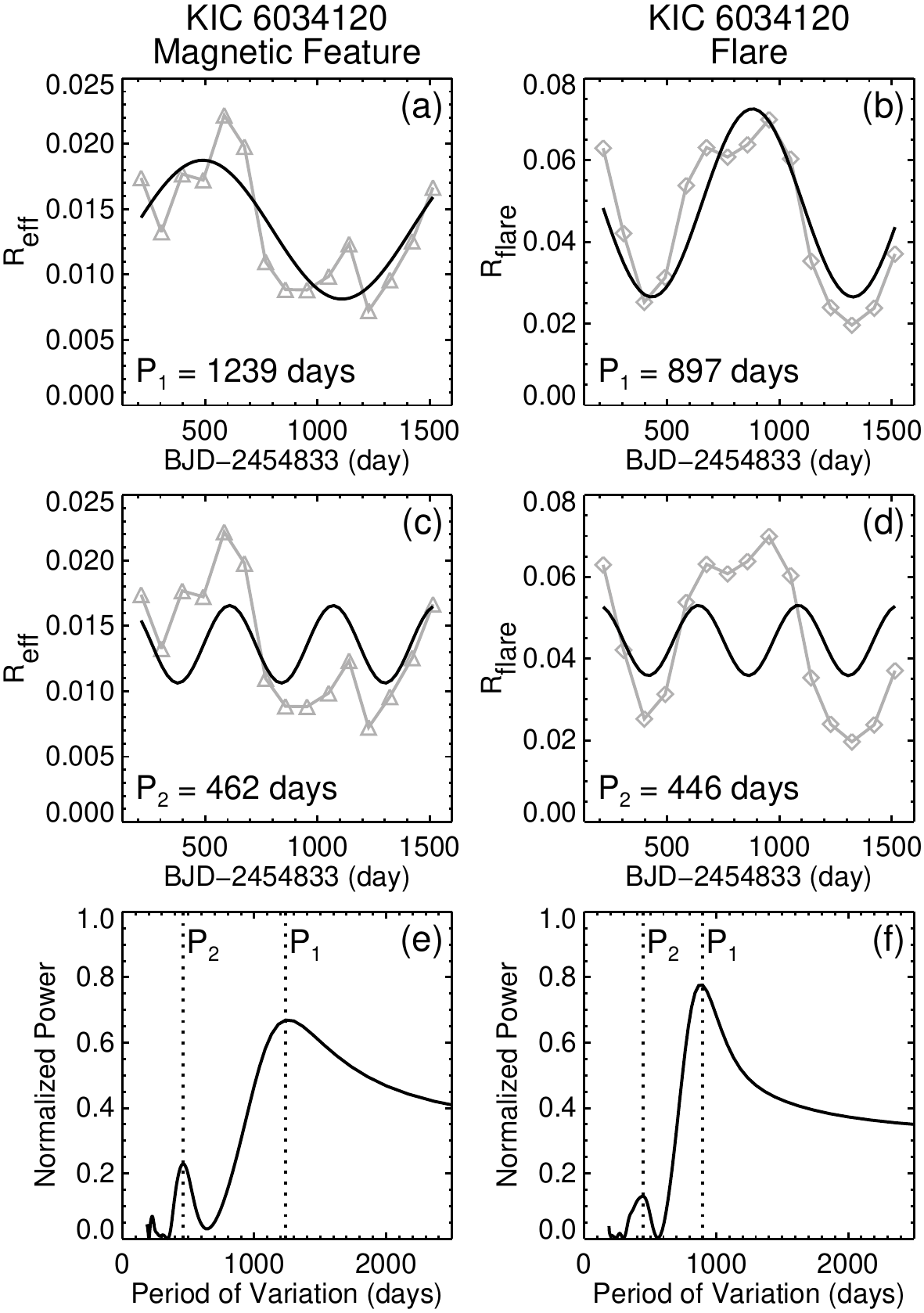}
  \caption{Evaluating the periods of the cyclic variations of $R_{\rm eff}$ and $R_{\rm flare}$ for the star KIC 6034120 using the GLS periodogram method. The left column is for $R_{\rm eff}$ (magnetic feature activity) and the right column is for $R_{\rm flare}$ (flare activity). The top row shows the first harmonic (fundamental) components of the $R_{\rm eff}$ and $R_{\rm flare}$ curves, and the middle row shows the second harmonic components. The harmonic components are plotted in black and the original $R_{\rm eff}$ and $R_{\rm flare}$ curves are in gray. The bottom row shows the GLS periodograms for the $R_{\rm eff}$ and $R_{\rm flare}$ curves. The two dotted lines in each periodogram indicate the first harmonic period $P_1$ and the second harmonic period $P_2$, which correspond to the two most prominent peaks in the periodogram. \label{fig10}}
\end{figure}

We quantitatively evaluate the cycle lengths (periods of the cyclic variations, denoted by $P_{\rm cyc}$) of $R_{\rm eff}$ and $R_{\rm flare}$ for the star KIC 6034120 using the generalized Lomb--Scargle (GLS) periodogram method \citep{2009A&A...496..577Z}. The GLS periodograms for the $R_{\rm eff}$ and $R_{\rm flare}$ curves of KIC 6034120 are shown in the bottom row of Figure \ref{fig10}; the left column is for $R_{\rm eff}$ (representing magnetic feature activity) and the right column is for $R_{\rm flare}$ (representing flare activity). In the top row of Figure \ref{fig10}, we show the first harmonic (fundamental) components of the $R_{\rm eff}$ and $R_{\rm flare}$ curves; and in the middle row we show the second harmonic components. The curves of the harmonic components are plotted in black and the original $R_{\rm eff}$ and $R_{\rm flare}$ curves are in gray. The values of the first harmonic period (denoted by $P_1$) and the second harmonic period (denoted by $P_2$) are also given in the corresponding panels in Figure \ref{fig10}. The first and second harmonic periods of $R_{\rm eff}$ or $R_{\rm flare}$ are evaluated through the two most prominent peaks in the associated periodograms (see the bottom row of Figure \ref{fig10}). The two periods (as well as the two peaks) are indicated by two vertical dotted lines in each periodogram and are labeled $P_1$ and $P_2$, respectively. The signals of higher-order harmonics are too low to be utilized.

The value of $P_{\rm cyc}$ can be evaluated either via the first harmonic period $P_1$ ($P_{\rm cyc}=P_1$) or via the second harmonic period $P_2$ ($P_{\rm cyc}=2 P_2$). The evaluated cycle length values of the magnetic feature activity (denoted by $P_{\rm cyc-M}$) through the parameter $R_{\rm eff}$ and the cycle length values of flare activity (denoted by $P_{\rm cyc-F}$) through the parameter $R_{\rm flare}$ for the star KIC 6034120 are summarized in Table \ref{tab:cycle-lengths}. It can been seen from Table \ref{tab:cycle-lengths} that the two values of $P_{\rm cyc-F}$ (flare activity) via the first harmonic and the second harmonic are almost identical (897 vs. 892 days), while the discrepancy between the two values of $P_{\rm cyc-M}$ (magnetic feature activity) is larger (1239 vs. 924 days). By comparing the FWHMs of the two peaks associated with the first and second harmonics in the periodogram of magnetic feature activity (see Figure \ref{fig10}(e)), the $P_{\rm cyc-M}$ value via the second harmonic (i.e., 924 days) is associated with a smaller FWHM and thus is more accurate (less uncertainty) than the value via the first harmonic. After all, the total time of the \emph{Kepler} observations from Q2 to Q16 (about 1388 days) only covers one full length of $P_1$ (see Figure \ref{fig10}(a)) but three full lengths of $P_2$ (see Figure \ref{fig10}(c)). That is why the FWHM of the peak associated with $P_1$ is larger in the periodogram. A similar property also exists for the periodogram of flare activity (see the right column of Figure \ref{fig10}).

\floattable
\begin{deluxetable}{lcccccr}
\tablecaption{Evaluated Cycle Lengths of Magnetic Feature Activity and Flare Activity for the Star KIC 6034120 \label{tab:cycle-lengths}}
\tablehead{
 && \colhead{Magnetic Feature Activity} && \colhead{Flare Activity}            && \colhead{Relative Difference\tablenotemark{a}} \\
 && \colhead{(through $R_{\rm eff}$)}   && \colhead{(through $R_{\rm flare}$)} &&                               \\
 && \colhead{$P_{\rm cyc-M}$ (days)}    && \colhead{$P_{\rm cyc-F}$ (days)}    &&
}
\startdata
via first harmonic ($P_{\rm cyc}=P_1$)     &&  1239  &&  897  &&  32.0\% \\
via second harmonic ($P_{\rm cyc}=2 P_2$)  &&  924   &&  892  &&  3.5\%  \\
\enddata
\tablenotetext{a}{Relative difference is defined as $\Delta/{\rm mean}$, where $\Delta$ is the absolute difference between the values.}
\end{deluxetable}

Based on the above analysis, we adopt the $P_{\rm cyc-M}$ and $P_{\rm cyc-F}$ values evaluated via the second harmonics of the $R_{\rm eff}$ and $R_{\rm flare}$ curves as the representative cycle lengths for the magnetic feature and flare activities of the star KIC 6034120, i.e., 924 days for $P_{\rm cyc-M}$ and 892 days for $P_{\rm cyc-F}$ (bottom line of Table \ref{tab:cycle-lengths}). The relative difference (see the rightmost column of Table \ref{tab:cycle-lengths}) between the two values is only 3.5\%. This result suggests that the long-term variations of the magnetic feature activity and the flare activity of the star KIC 6034120 has a similar cycle length. Together with the fact that the two cyclic variations are not in phase with each other (see Section \ref{subsec:corr}), this might indicate that the two activities are controlled by two distinct aspects of the magnetic field generated through the same dynamo process.

\section{Summary and Discussion} \label{sec:sum}

In this paper, we analyze the light curve data of three solar-type stars (KIC 6034120, KIC 3118883, and KIC 10528093) observed with \emph{Kepler} and investigate the relationship between their magnetic feature activities and flare activities. The information on magnetic feature activity is deduced from the rotational modulation signal in the light curves and is quantitatively described by two measures (magnetic proxies) of the light curves suggested in Paper I. The first measure, $i_{\rm AC}$, describes the degree of periodicity of the light curves, which reflects the stability of the magnetic features that cause the rotational modulation. The second measure, $R_{\rm eff}$, describes the effective range of light curve fluctuation, which reflects the spatial size or coverage of magnetic features. The analysis shows that the time variations of the two measures are positively correlated for all three investigated stars. In this work, we adopt $R_{\rm eff}$ as the representative quantity of magnetic feature activities.

The information on flare activity is deduced from the flare spike signals in the light curves. We employ three flare indexes to quantitatively describe the flare activity properties, which are the time occupation ratio of flares $R_{\rm flare}$ (equation (\ref{equ:Rflare})), total relative power (energy rate) of flares $P_{\rm flare}$ (equation (\ref{equ:Pflare})), and averaged relative flux magnitude of flares $M_{\rm flare}$ (equation (\ref{equ:Mflare})). These flare indexes reflect different aspects (frequency, energy, and magnitude) of flare activity properties and are connected with each other by equation (\ref{equ:RPMflare}). The analysis shows that for the three investigated stars, $R_{\rm flare}$ and $P_{\rm flare}$ have similar time variations, while $M_{\rm flare}$ fluctuates randomly around its mean value. In this work, we adopt $R_{\rm flare}$ as the representative quantity of flare activity.

By analyzing the relationship between the two parameters $R_{\rm eff}$ and $R_{\rm flare}$ quantitatively, we found that: (1) both the magnetic feature activity and the flare activity exhibit long-term variations as the Sun does; (2) unlike the Sun, the long-term variations of magnetic feature activity and flare activity are not in phase with each other; (3) the analysis of the star KIC 6034120, which has the highest flare occurrence frequency among the three stars, suggests that the long-term variations of the magnetic feature activity and flare activity have a similar cycle length.

Our analysis and results indicate that the magnetic features that dominate rotational modulation and the flares possibly have different source regions, although they may be influenced by the magnetic field generated through the same dynamo process.

We speculate that the positive correlation between the $R_{\rm eff}$ and $i_{\rm AC}$ suggests that the magnetic field associated with rotational modulation on these three stars, unlike the Sun, is an agent of stability more than activity. We further speculate that the poloidal components of the magnetic field on these stars are stronger than the toroidal components of the magnetic field \citep[for a brief description of the poloidal and toroidal magnetic fields in the context of the stellar dynamo model, see, e.g.,][]{2017SCPMA..6019601C}. According to \citet{2006ApJ...644..575Z}, the poloidal component of the magnetic field is an agent for confinement and stability and the torodial component of the magnetic field is an agent for magnetic free energy and eruption; the stronger component of the poloidal field naturally explained why the $R_{\rm eff}$ and $i_{\rm AC}$ of our results are positively correlated and why the $R_{\rm eff}$  and $R_{\rm flare}$ are not related or even negatively related, contrary to what happens on the Sun.

\acknowledgments
This paper includes data collected by the \emph{Kepler} mission. Funding for the \emph{Kepler} mission is provided by the NASA Science Mission directorate. The \emph{Kepler} data presented in this paper (\emph{Kepler} Data Release 25) were obtained from the Mikulski Archive for Space Telescopes (MAST). This work is jointly supported by the National Natural Science Foundation of China (NSFC) through grants 11761141002, 11403044, 11603040, 11473040, 40890160, 40890161, 10803011, 11221063, 11303051, 11125314, 11273031, and U1531247; the Strategic Priority Research Program on Space Science, Chinese Academy of Sciences (grant XDA04060801); the National Science and Technology Basic Work Program of China (grant 2014FY120300); the National Basic Research Program of China (973 Program) through grant 2011CB811406, and the China Meteorological Administration (grant GYHY201106011).

\appendix

\section{Evaluating Noise Levels of the Light Curve Data} \label{sec:appendix_noise_level}

\begin{figure}
  \epsscale{1.0}
  \plotone{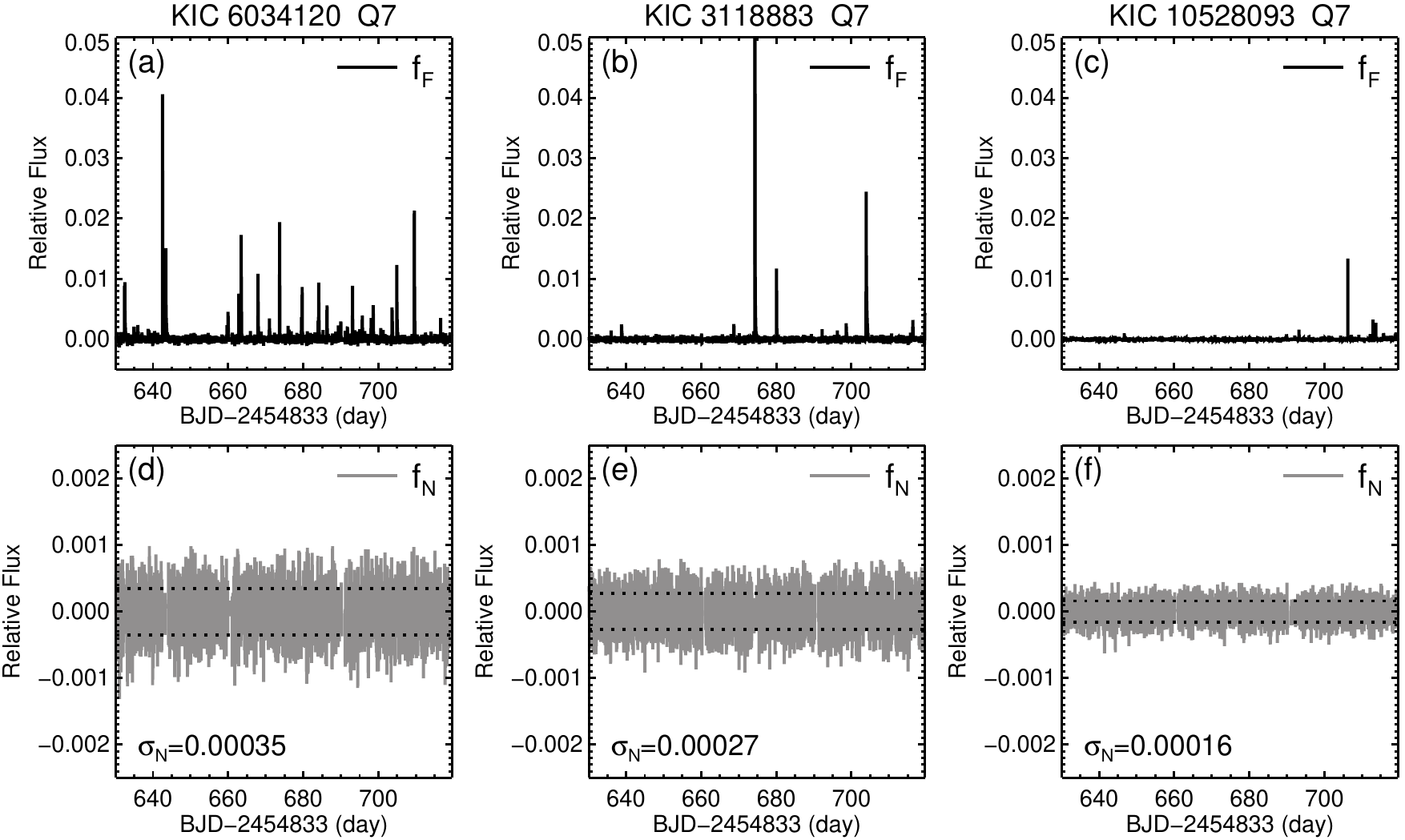}
  \caption{Evaluation of the noise levels ($\sigma_{\rm N}$) of the light curve data for the three investigated stars. Each column corresponds to one star. (a)--(c) The $f_{\rm F}$ (flare component) curves of the three stars in Q7. (d)--(f)The $f_{\rm N}$ (background noise component) curves of the three stars in Q7 and the evaluated $\sigma_{\rm N}$ values. The two horizontal dotted lines in each plot of $f_{\rm N}$ indicate the positions of $\pm \sigma_{\rm N}$.  \label{fig11}}
\end{figure}

In Section \ref{sec:data}, the original \emph{Kepler} light curve data are separated into the gradual component ($f_{\rm G}$) and flare component ($f_{\rm F}$), and the noises in the original light curves are left in the $f_{\rm F}$ data. So it is convenient to utilize the $f_{\rm F}$ data for noise level evaluation. Since the photometric precision of \emph{Kepler} for a given star is stable across the quarters (see Figures \ref{fig3}(c), \ref{fig4}(c), and \ref{fig5}(c)), we only use the $f_{\rm F}$ data in Q7 to evaluate the light curve noise levels for the three investigated stars. The top row of Figure \ref{fig11} displays the $f_{\rm F}$ curves of the three stars in Q7. We manually removed all the flare spikes and the time gaps (for spacecraft data downlink, see Figure \ref{fig6}) from the $f_{\rm F}$ flux data, and obtained the flux data of pure background noises (denoted by $f_{\rm N}$, where the letter `N' refers to `noise'). The $f_{\rm N}$ curves of the three stars in Q7 are displayed in the bottom row of Figure \ref{fig11}. The noise levels (denoted by $\sigma_{\rm N}$) of the light curve data are quantitatively measured by the standard deviation of the $f_{\rm N}$ data:
\begin{equation}
  \sigma_{\rm N}=\sqrt{\langle {{f_{\rm N}}^2}\rangle}.
\end{equation}

As given in the bottom row of Figure \ref{fig11}, the derived noise level ($\sigma_{\rm N}$) values for the three investigated stars (KIC 6034120, KIC 3118883, and KIC 10528093) are 0.00035, 0.00027, and 0.00016, respectively, which are also illustrated in the bottom row of Figures \ref{fig11} by two horizontal dotted lines (indicating the positions of $\pm \sigma_{\rm N}$) in each panel. Note that these noise level values are consistent with the \emph{Kepler} magnitude ($Kp$) values (see Table \ref{tab:star-param}) of the three stars, that is, the stars with smaller $Kp$ values (brighter stars) have lower noise levels \citep{2011ApJS..197....6G, 2015AJ....150..133G}.

\section{Error Estimates on $i_{\rm AC}$ and $R_{\rm eff}$} \label{sec:appendix_error_estimates}

\subsection{Equations to Calculate the Two Parameters} \label{subsec:parameters_equ}

For a time-series of flux data with $N$ evenly spaced data points $\{X_t, t=0, 1, \ldots, N-1\}$, the parameter $i_{\rm AC}$ (autocorrelation index) is calculated by equations
\begin{equation}\label{equ:acf}
  \rho(h)=\frac{\sum_{t=0}^{N-1-h} (X_{t+h}-\overline{X})(X_t-\overline{X})}{\sum_{t=0}^{N-1} (X_t-\overline{X})^2},
  \qquad 0\leqslant h \leqslant N-1,
\end{equation}
\begin{equation}\label{equ:iac}
  i_{\rm AC}=\frac{2}{N} \int_{0}^{N/2} |\rho(h)| dh,
\end{equation}
where $\overline{X}$ is the mean value of the time-series $\{X_t\}$, $\rho(h)$ is the autocorrelation coefficient of $\{X_t\}$ at time lag $h$, and $i_{\rm AC}$ is defined as the average value of $|\rho(h)|$ for the first half of the $\rho(h)$ function. The value of $i_{\rm AC}$ is in the interval between 0 and $\frac{3}{2\pi}$ ($\approx0.477$) and larger $i_{\rm AC}$ means stronger periodicity (see Paper I for more details).

The parameter $R_{\rm eff}$ (effective range of fluctuation) for the time-series $\{X_t\}$ is calculated by equations
\begin{equation}\label{equ:x_t}
  x_t=\frac{X_t-\widetilde{X}}{\widetilde{X}},
  \qquad t=0, 1, \ldots, N-1,
\end{equation}
\begin{equation} \label{equ:reff}
  R_{\rm eff}=2\sqrt{2} \cdot x_{\rm rms}=2\sqrt{2} \cdot \sqrt{\frac{1}{N} \sum_{t=0}^{N-1} x_t^2},
\end{equation}
where $\widetilde{X}$ is the median value of $\{X_t\}$, $\{x_t, t=0, 1, \ldots, N-1\}$ is the relative flux expression of $\{X_t\}$,  and $x_{\rm rms}$ is the rms value of $\{x_t\}$. In equation (\ref{equ:reff}), the rms value is multiplied by a scaling factor $2\sqrt{2}$ to obtain the effective range between crest and trough (see Paper I for more details).

\subsection{Errors Caused by the Photometric Noises}
Since we use the $f_{\rm M}$ (rotational modulation component) data to calculate $i_{\rm AC}$ and $R_{\rm eff}$ (see Section \ref{subsec:proxies}) and the background photometric noises have been filtered out from the $f_{\rm G}$ and hence the $f_{\rm M}$ flux data (see Section \ref{subsec:f_M}), the errors caused by the photometric noises (see Appendix \ref{sec:appendix_noise_level}) on the two parameters can be omitted.

\subsection{Errors Caused by the Cutoff Effect}

\begin{figure}
  \epsscale{0.37}
  \plotone{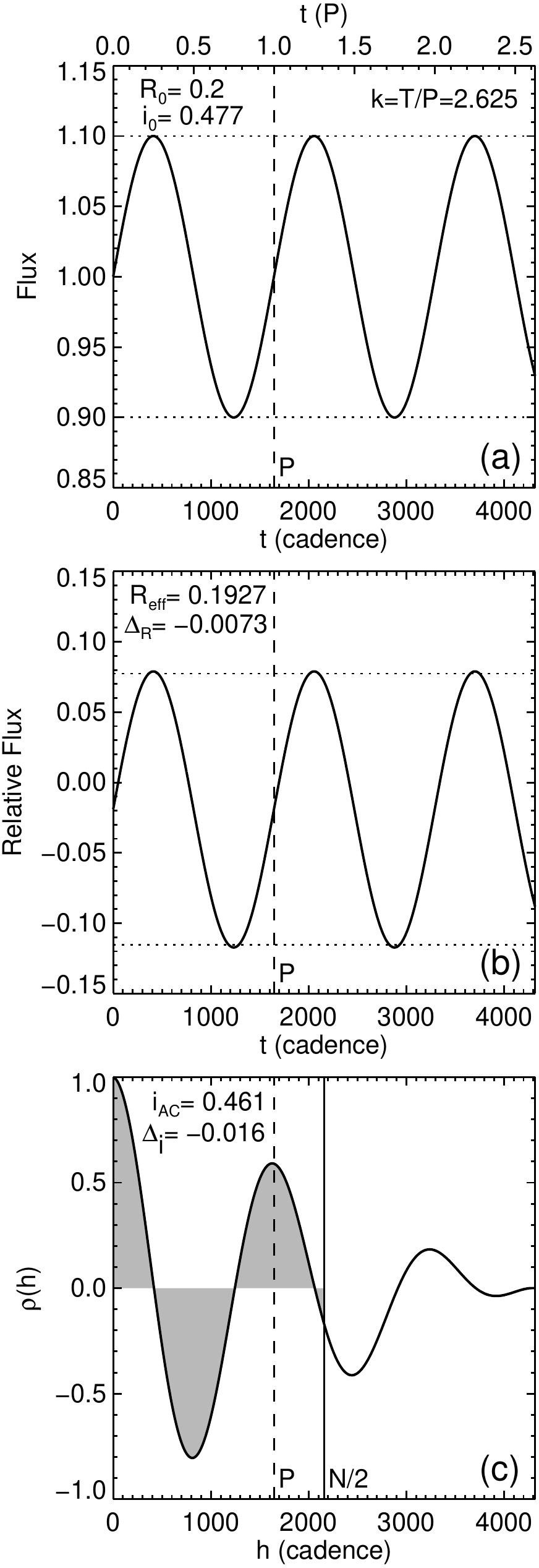}
  \caption{(a) An example of a sine curve with a cutoff end. $k$ is the period number of the curve. $R_0$ and $i_0$ give the standard values of $R_{\rm eff}$ and $i_{\rm AC}$ for the sine curve (see the main text). The two horizontal dotted lines indicate the distance between the crest and trough of the sine curve. (b) Relative flux and the computed $R_{\rm eff}$ of the sine curve. $\Delta_R=R_{\rm eff}-R_0$. The distance between the two horizontal dotted lines illustrates the computed value of $R_{\rm eff}$. (c) Plot of the $\rho(h)$ function and the computed $i_{\rm AC}$ of the sine curve. $\Delta_i=i_{\rm AC}-i_0$. The shaded areas and the vertical solid line (labeled with `N/2') indicate the first half of the $\rho(h)$ function employed for evaluating $i_{\rm AC}$ (see the main text). The vertical dashed lines (labeled with `P') in the panels indicate the position of one period. \label{fig12}}
\end{figure}

The two parameters $i_{\rm AC}$ and $R_{\rm eff}$ are defined based on the rotational modulation property of light curves. For a practical light curve, the periodic fluctuation will be cut off at the two ends of the light curve, which may introduce extra uncertainties (errors) on the two parameters.

We use light curves simulated from a sine function to demonstrate the cutoff effect on the values of $i_{\rm AC}$ and $R_{\rm eff}$, and then estimate the errors caused by this effect. A general sine light curve can be expressed by
\begin{equation}\label{equ:func_sine_curve}
  F=F_0\left[A\sin\left(2\pi\frac{t}{P}+\varphi\right)+1\right],
\end{equation}
where $F_0$ is the baseline of the flux, $A$ is the relative amplitude of the light curve fluctuation, $P$ is the period, and $\varphi$ is the initial phase. Figure \ref{fig12}(a) gives an example of a sine curve with a cutoff end. Here, without loss of generality, we adopt $F_0$ as unity (i.e., $F_0$=1) and $A=0.1$ as shown in Figure \ref{fig12}(a). The initial phase $\varphi$ of the example curve is $0$. We define the period number $k$ of a periodic light curve as
\begin{equation}\label{equ:definition_k}
  k=T/P,
\end{equation}
where $T$ is the total time length of the light curve. For the example light curve, $k=2.625$, as shown in Figure \ref{fig12}(a). If $k$ is sufficiently large, the values of $i_{\rm AC}$ and $R_{\rm eff}$ for a sine curve approach $\frac{3}{2\pi}\approx 0.477$ and $2A$ ($=0.2$ for this example), respectively (see explanation in Paper I, and diagram illustration below). We call them the standard values of $i_{\rm AC}$ and $R_{\rm eff}$ for the sine curve and denote them by $i_0$ and $R_0$. If $k$ is a finite number, as for the example curve, the computed values of $i_{\rm AC}$ and $R_{\rm eff}$ will deviate from $i_0$ and $R_0$. Figure \ref{fig12}(b) shows the plot of the relative flux (computed by equation (\ref{equ:x_t})), and Figure \ref{fig12}(c) shows the plot of the $\rho(h)$ function (defined in equation (\ref{equ:acf})) of the example light curve. The computed values of $i_{\rm AC}$ and $R_{\rm eff}$ from equations (\ref{equ:iac}) and (\ref{equ:reff}), as well as their deviations from the standard values (denoted by $\Delta_i=i_{\rm AC}-i_0$ and $\Delta_R=R_{\rm eff}-R_0$), are given in Figure \ref{fig12}.

\begin{figure}
  \epsscale{0.71}
  \plotone{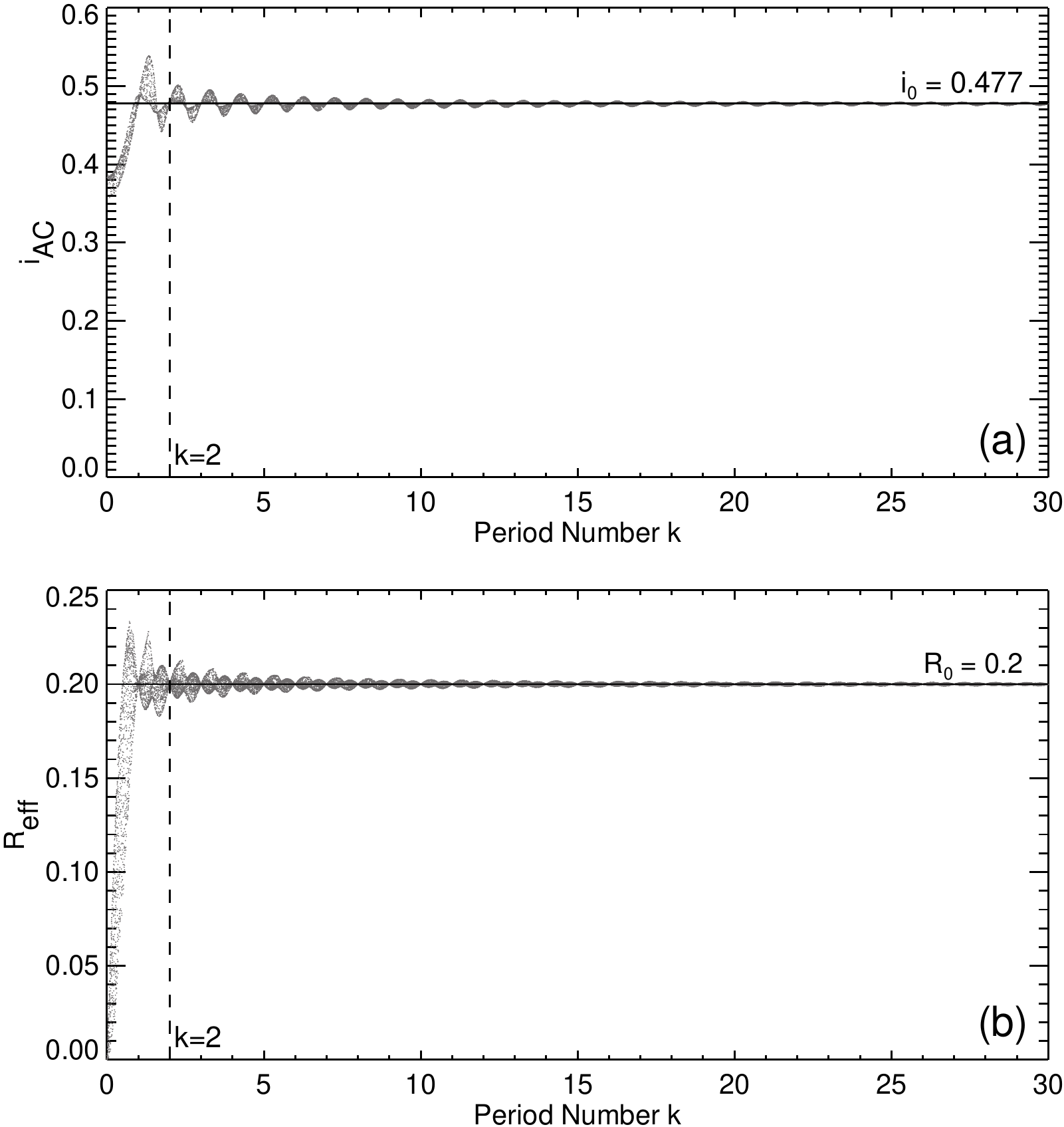}
  \caption{Scatter plots of $i_{\rm AC}$ and $R_{\rm eff}$ versus period number $k$ after the Monte Carlo simulation in the parameter space $\{(\varphi, k)\}$. Each data point corresponds to a randomly generated $(\varphi, k)$ pair (see the main text). The horizontal solid lines indicate the standard values of the two parameters, $i_0=0.477$ and $R_0=0.2$. The vertical dashed line in each panel marks the position of $k=2$ (see the main text). \label{fig13}}
\end{figure}

Besides the period number $k$, the initial phase ($\varphi$ in equation (\ref{equ:func_sine_curve})) of a sine curve can also affect the values of $i_{\rm AC}$ and $R_{\rm eff}$. For the example sine curve shown in Figure \ref{fig12}(a), $\varphi=0$; while for a general sine curve, $\varphi$ can vary between $0$ to $2\pi$. To find out the distributions of $i_{\rm AC}$ and $R_{\rm eff}$ values for different combinations of $\varphi$ and $k$, we made a Monte Carlo simulation in the parameter space $\{(\varphi, k)\}$ where $\varphi \in (0, 2\pi)$ and $k \in (0, 30)$. For each randomly generated $(\varphi, k)$ pair, we calculated the corresponding values of $i_{\rm AC}$ and $R_{\rm eff}$. The resultant scatter plots of $i_{\rm AC}$ and $R_{\rm eff}$ versus period number $k$ are shown in Figure \ref{fig13}.

It can be seen from Figure \ref{fig13} that the computed values of $i_{\rm AC}$ and $R_{\rm eff}$ fluctuate around their standard values $i_0=0.477$ and $R_0=0.2$. With the increase of $k$ (i.e., more periods being included in light curves), the fluctuation amplitude decreases and the values of the two parameters approach $i_0$ and $R_0$.

The deviations of $i_{\rm AC}$ and $R_{\rm eff}$ from their standard values are the consequence of the cutoff effect. Thus, the uncertainties (errors) of the two parameters caused by the cutoff effect can be reflected by the amplitudes of fluctuation shown in Figure \ref{fig13}. The larger the $k$ is, the smaller the error is. On the other hand, when $k$ is less than 2, the errors of the two parameters are extremely large and the results of $i_{\rm AC}$ and $R_{\rm eff}$ are unreliable. For practical usage of the two metrics, at least two periods should be included in a light curve.

\begin{figure}
  \epsscale{0.72}
  \plotone{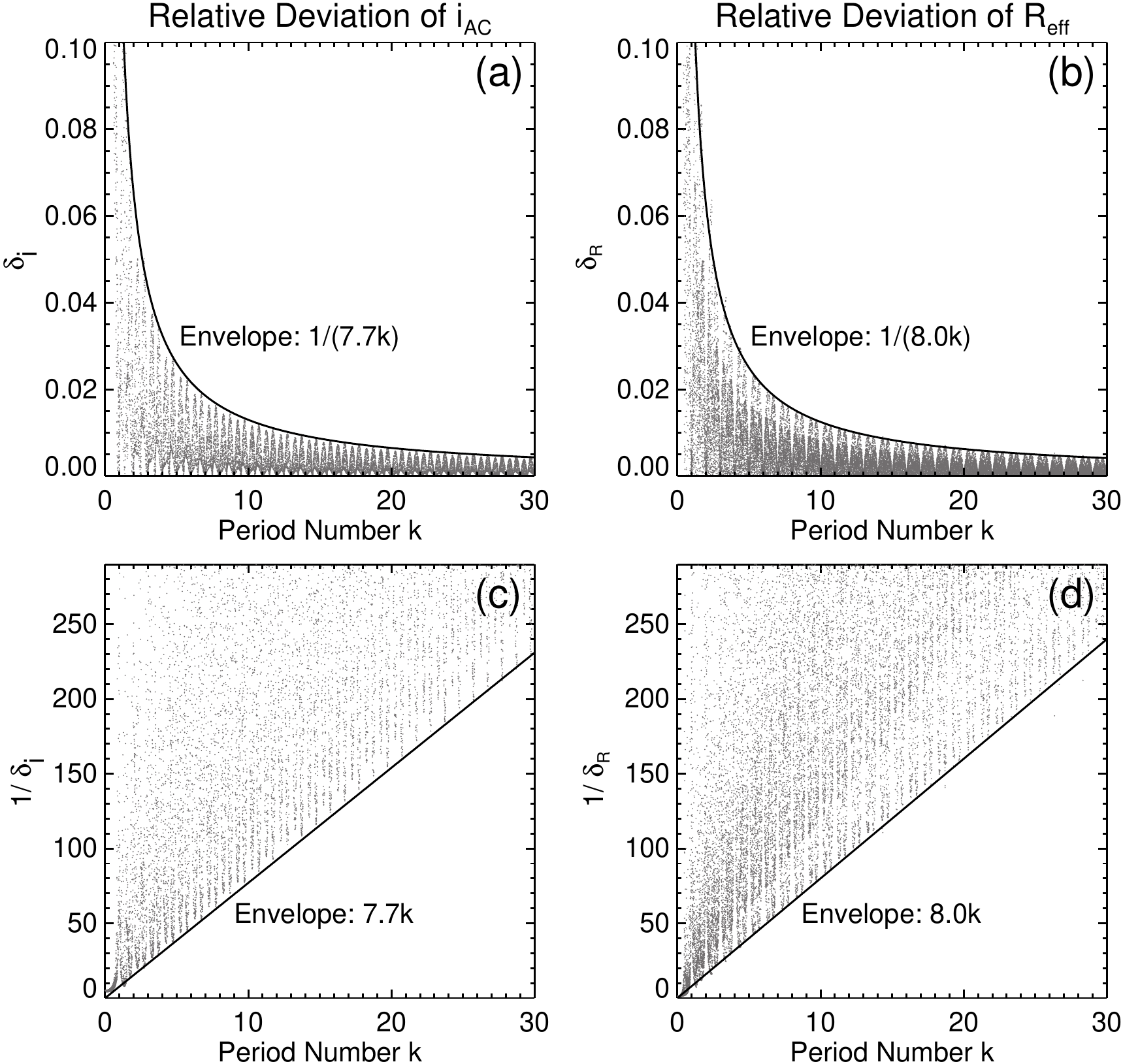}
  \caption{(a)--(b) Scatter plots of $\delta_i$ and $\delta_R$ (relative deviations of $i_{\rm AC}$ and $R_{\rm eff}$) versus period number $k$. The solid curves (reciprocal functions of $k$) fit the upper envelopes of the scatter plots. (c)--(d) Scatter plots of $1/\delta_i$ and $1/\delta_R$ versus $k$. The solid lines (linear functions of $k$) fit the lower envelopes of the scatter plots. \label{fig14}}
\end{figure}

To evaluate the errors caused by the cutoff effect quantitatively, we calculated the relative deviations of $i_{\rm AC}$ and $R_{\rm eff}$ from their standard values for all the data points shown in Figure \ref{fig13} by equations
\begin{equation}\label{equ:rela_delta_i}
  \delta_i=|\Delta_i|/i_0=|i_{\rm AC}-i_0|/i_0,
\end{equation}
\begin{equation}\label{equ:rela_delta_R}
  \delta_R=|\Delta_R|/R_0=|R_{\rm eff}-R_0|/R_0,
\end{equation}
where $|\Delta_i|$ and $|\Delta_R|$ are absolute deviations, and $\delta_i$ and $\delta_R$ are relative deviations. The scatter plots of $\delta_i$ and $\delta_R$ versus period number $k$ are displayed in Figures \ref{fig14}(a) and (b). It can be seen that the distribution ranges of $\delta_i$ and $\delta_R$ decrease with the increase of $k$. We evaluate the relative errors caused by the cutoff effect on the two parameters by the heights of the upper envelopes of the scatter plots in Figures \ref{fig14}(a) and (b).

In Figures \ref{fig14}(c) and (d), we show the scatter plots of $1/\delta_i$ and $1/\delta_R$ versus $k$. It can be seen that the lower envelopes of $1/\delta_i$ and $1/\delta_R$ can be empirically fitted by linear functions of $k$ ($7.7k$ for $1/\delta_i$ and $8.0k$ for $1/\delta_R$, illustrated with solid lines in Figures \ref{fig14}(c) and (d)). Then, the upper envelopes of $\delta_i$ and $\delta_R$ in Figures \ref{fig14}(a) and (b) (and hence the relative errors of $i_{\rm AC}$ and $R_{\rm eff}$) can be empirically fitted by reciprocal functions of $k$ ($\frac{1}{7.7k}$ for $\delta_i$ and $\frac{1}{8.0k}$ for $\delta_R$, illustrated with solid curves in Figures \ref{fig14}(a) and (b)).

The absolute errors of $i_{\rm AC}$ and $R_{\rm eff}$ (denoted by $\epsilon_i$ and $\epsilon_R$) can be calculated by equations
\begin{equation}\label{equ:error_iac}
  \epsilon_i=\frac{1}{7.7\times k} i_{\rm AC},
\end{equation}
\begin{equation}\label{equ:error_reff}
  \epsilon_R\approx \frac{1}{8.0\times k} R_{\rm eff},
\end{equation}
where the relative errors are multiplied by the actual values of $i_{\rm AC}$ and $R_{\rm eff}$ to obtain the absolute errors.

In Section \ref{subsec:proxies}, the errors of $i_{\rm AC}$ and $R_{\rm eff}$ in each quarter for the three investigated stars (see Table \ref{tab:iac-reff}) are calculated based on equations (\ref{equ:error_iac}) and (\ref{equ:error_reff}). We use the nominal values of $P_{\rm rot}$ listed in Table \ref{tab:star-param} to calculate the period number $k=T_q/P_{\rm rot}$ covered by each quarter, where $T_q$ is the time length of the quarter (time distance between the first and last cadences). Note that the actual errors are influenced by two factors: the time length of the quarter and the absolute values of the two parameters. That is, the longer the quarter is, the smaller the errors are; if the absolute values of the two parameters are larger, the corresponding errors are also larger.

\section{Error Estimates on $R_{\rm flare}$, $P_{\rm flare}$, and $M_{\rm flare}$} \label{sec:appendix_error_estimates_flare}

To evaluate the errors of the three flare indexes $R_{\rm flare}$, $P_{\rm flare}$, and $M_{\rm flare}$ employed in Section \ref{subsec:indexes}, we rewrite $T_{\rm flare}$ and $T_{\rm obs}$ (see the definitions in Section \ref{subsec:indexes}) in the form of cadences:
\begin{equation}\label{equ:tflare_cadence}
  T_{\rm flare}= m,
\end{equation}
\begin{equation}\label{equ:tobs_cadence}
  T_{\rm obs}= n,
\end{equation}
where $m$ and $n$ are the cadence numbers covered by $T_{\rm flare}$ and $T_{\rm obs}$. Then, equations (\ref{equ:Rflare}), (\ref{equ:Pflare}), and (\ref{equ:Mflare}) of $R_{\rm flare}$, $P_{\rm flare}$, and $M_{\rm flare}$ in Section \ref{subsec:indexes} can be rewritten as
\begin{equation}\label{equ:Rflare_cadence}
  R_{\rm flare}=\frac{m}{n},
\end{equation}
\begin{equation}\label{equ:Pflare_cadence}
  P_{\rm flare}=\frac{\sum_{t=0}^{m-1}f_{\rm S}(t)}{n},
\end{equation}
\begin{equation}\label{equ:Mflare_cadence}
  M_{\rm flare}=\frac{\sum_{t=0}^{m-1}f_{\rm S}(t)}{m}.
\end{equation}

Because the precision of the measurement of $m$ is one cadence, the error of $R_{\rm flare}$ (denoted by $\epsilon^{R}_{\rm flare}$) can be calculated by equation
\begin{equation}\label{equ:e_rflare}
  \epsilon^{R}_{\rm flare}=\frac{1}{n}.
\end{equation}
The noise levels $\sigma_{\rm N}$ of the $f_{\rm F}$ flux data (and hence the $f_{\rm S}$ data) have been evaluated for the three investigated stars in Appendix \ref{sec:appendix_noise_level}. Then, the errors of $P_{\rm flare}$ and $M_{\rm flare}$ (denoted by $\epsilon^{P}_{\rm flare}$ and $\epsilon^{M}_{\rm flare}$) can be calculated by equations
\begin{equation}\label{equ:e_pflare}
  \epsilon^{P}_{\rm flare}=\frac{\sqrt{m}\cdot\sigma_{\rm N}}{n},
\end{equation}
\begin{equation}\label{equ:e_mflare}
  \epsilon^{M}_{\rm flare}=\frac{\sigma_{\rm N}}{\sqrt{m}}.
\end{equation}

In Section \ref{subsec:indexes}, the errors of the three flare indexes $\epsilon^{R}_{\rm flare}$, $\epsilon^{P}_{\rm flare}$, and $\epsilon^{M}_{\rm flare}$ in each quarter for the three investigated stars (see Table \ref{tab:flare-indexes}) are calculated based on equations (\ref{equ:e_rflare})--(\ref{equ:e_mflare}).


\begin{thebibliography}{}
\expandafter\ifx\csname natexlab\endcsname\relax\def\natexlab#1{#1}\fi

\bibitem[{{Aschwanden} \& {Freeland}(2012)}]{2012ApJ...754..112A}
{Aschwanden}, M.~J., \& {Freeland}, S.~L. 2012, \apj, 754, 112

\bibitem[{{Auvergne} {et~al.}(2009){Auvergne}, {Bodin}, {Boisnard}, {Buey},
  {Chaintreuil}, {Epstein}, {Jouret}, {Lam-Trong}, {Levacher}, {Magnan},
  {Perez}, {Plasson}, {Plesseria}, {Peter}, {Steller}, {Tiph{\`e}ne}, {Baglin},
  {Agogu{\'e}}, {Appourchaux}, {Barbet}, {Beaufort}, {Bellenger}, {Berlin},
  {Bernardi}, {Blouin}, {Boumier}, {Bonneau}, {Briet}, {Butler}, {Cautain},
  {Chiavassa}, {Costes}, {Cuvilho}, {Cunha-Parro}, {de Oliveira Fialho},
  {Decaudin}, {Defise}, {Djalal}, {Docclo}, {Drummond}, {Dupuis}, {Exil},
  {Faur{\'e}}, {Gaboriaud}, {Gamet}, {Gavalda}, {Grolleau}, {Gueguen},
  {Guivarc'h}, {Guterman}, {Hasiba}, {Huntzinger}, {Hustaix}, {Imbert},
  {Jeanville}, {Johlander}, {Jorda}, {Journoud}, {Karioty}, {Kerjean},
  {Lafond}, {Lapeyrere}, {Landiech}, {Larqu{\'e}}, {Laudet}, {Le Merrer},
  {Leporati}, {Leruyet}, {Levieuge}, {Llebaria}, {Martin}, {Mazy}, {Mesnager},
  {Michel}, {Moalic}, {Monjoin}, {Naudet}, {Neukirchner}, {Nguyen-Kim},
  {Ollivier}, {Orcesi}, {Ottacher}, {Oulali}, {Parisot}, {Perruchot},
  {Piacentino}, {Pinheiro da Silva}, {Platzer}, {Pontet}, {Pradines},
  {Quentin}, {Rohbeck}, {Rolland}, {Rollenhagen}, {Romagnan}, {Russ}, {Samadi},
  {Schmidt}, {Schwartz}, {Sebbag}, {Smit}, {Sunter}, {Tello}, {Toulouse},
  {Ulmer}, {Vandermarcq}, {Vergnault}, {Wallner}, {Waultier}, \&
  {Zanatta}}]{2009A&A...506..411A}
{Auvergne}, M., {Bodin}, P., {Boisnard}, L., {et~al.} 2009, \aap, 506, 411

\bibitem[{{Baliunas} {et~al.}(1983){Baliunas}, {Hartmann}, {Noyes}, {Vaughan},
  {Preston}, {Frazer}, {Lanning}, {Middelkoop}, \&
  {Mihalas}}]{1983ApJ...275..752B}
{Baliunas}, S.~L., {Hartmann}, L., {Noyes}, R.~W., {et~al.} 1983, \apj, 275,
  752

\bibitem[{{Balona}(2015)}]{2015MNRAS.447.2714B}
{Balona}, L.~A. 2015, \mnras, 447, 2714

\bibitem[{{Basri} {et~al.}(2013){Basri}, {Walkowicz}, \&
  {Reiners}}]{2013ApJ...769...37B}
{Basri}, G., {Walkowicz}, L.~M., \& {Reiners}, A. 2013, \apj, 769, 37

\bibitem[{{Basri} {et~al.}(2010){Basri}, {Walkowicz}, {Batalha}, {Gilliland},
  {Jenkins}, {Borucki}, {Koch}, {Caldwell}, {Dupree}, {Latham}, {Meibom},
  {Howell}, \& {Brown}}]{2010ApJ...713L.155B}
{Basri}, G., {Walkowicz}, L.~M., {Batalha}, N., {et~al.} 2010, \apjl, 713, L155

\bibitem[{{Basri} {et~al.}(2011){Basri}, {Walkowicz}, {Batalha}, {Gilliland},
  {Jenkins}, {Borucki}, {Koch}, {Caldwell}, {Dupree}, {Latham}, {Marcy},
  {Meibom}, \& {Brown}}]{2011AJ....141...20B}
---. 2011, \aj, 141, 20

\bibitem[{{Benz}(2008)}]{2008LRSP....5....1B}
{Benz}, A.~O. 2008, Living Reviews in Solar Physics, 5, 1

\bibitem[{{Borucki}(2016)}]{2016RPPh...79c6901B}
{Borucki}, W.~J. 2016, Reports on Progress in Physics, 79, 036901

\bibitem[{{Borucki} {et~al.}(2010){Borucki}, {Koch}, {Basri}, {Batalha},
  {Brown}, {Caldwell}, {Caldwell}, {Christensen-Dalsgaard}, {Cochran},
  {DeVore}, {Dunham}, {Dupree}, {Gautier}, {Geary}, {Gilliland}, {Gould},
  {Howell}, {Jenkins}, {Kondo}, {Latham}, {Marcy}, {Meibom}, {Kjeldsen},
  {Lissauer}, {Monet}, {Morrison}, {Sasselov}, {Tarter}, {Boss}, {Brownlee},
  {Owen}, {Buzasi}, {Charbonneau}, {Doyle}, {Fortney}, {Ford}, {Holman},
  {Seager}, {Steffen}, {Welsh}, {Rowe}, {Anderson}, {Buchhave}, {Ciardi},
  {Walkowicz}, {Sherry}, {Horch}, {Isaacson}, {Everett}, {Fischer}, {Torres},
  {Johnson}, {Endl}, {MacQueen}, {Bryson}, {Dotson}, {Haas}, {Kolodziejczak},
  {Van Cleve}, {Chandrasekaran}, {Twicken}, {Quintana}, {Clarke}, {Allen},
  {Li}, {Wu}, {Tenenbaum}, {Verner}, {Bruhweiler}, {Barnes}, \&
  {Prsa}}]{2010Sci...327..977B}
{Borucki}, W.~J., {Koch}, D., {Basri}, G., {et~al.} 2010, Science, 327, 977

\bibitem[{{Brown} {et~al.}(2011){Brown}, {Latham}, {Everett}, \&
  {Esquerdo}}]{2011AJ....142..112B}
{Brown}, T.~M., {Latham}, D.~W., {Everett}, M.~E., \& {Esquerdo}, G.~A. 2011,
  \aj, 142, 112

\bibitem[{{Chaplin} {et~al.}(2011){Chaplin}, {Bedding}, {Bonanno}, {Broomhall},
  {Garc{\'{\i}}a}, {Hekker}, {Huber}, {Verner}, {Basu}, {Elsworth}, {Houdek},
  {Mathur}, {Mosser}, {New}, {Stevens}, {Appourchaux}, {Karoff}, {Metcalfe},
  {Molenda-{\.Z}akowicz}, {Monteiro}, {Thompson}, {Christensen-Dalsgaard},
  {Gilliland}, {Kawaler}, {Kjeldsen}, {Ballot}, {Benomar}, {Corsaro},
  {Campante}, {Gaulme}, {Hale}, {Handberg}, {Jarvis}, {R{\'e}gulo}, {Roxburgh},
  {Salabert}, {Stello}, {Mullally}, {Li}, \& {Wohler}}]{2011ApJ...732L...5C}
{Chaplin}, W.~J., {Bedding}, T.~R., {Bonanno}, A., {et~al.} 2011, \apjl, 732,
  L5

\bibitem[{Chatfield(2003)}]{Chatfield-2003}
Chatfield, C. 2003, The analysis of time series: An introduction, 6th Edition
  (Chapman \& Hall)

\bibitem[{{Choudhuri}(2017)}]{2017SCPMA..6019601C}
{Choudhuri}, A.~R. 2017, Science China Physics, Mechanics, and Astronomy, 60,
  019601

\bibitem[{{Cui} {et~al.}(2012){Cui}, {Zhao}, {Chu}, {Li}, {Li}, {Zhang}, {Su},
  {Yao}, {Wang}, {Xing}, {Li}, {Zhu}, {Wang}, {Gu}, {Luo}, {Xu}, {Zhang},
  {Liu}, {Zhang}, {Yang}, {Cao}, {Chen}, {Chen}, {Chen}, {Chen}, {Chu}, {Feng},
  {Gong}, {Hou}, {Hu}, {Hu}, {Hu}, {Jia}, {Jiang}, {Jiang}, {Jiang}, {Jin},
  {Li}, {Li}, {Li}, {Liu}, {Liu}, {Lu}, {Mao}, {Men}, {Qi}, {Qi}, {Shi},
  {Tang}, {Tao}, {Wang}, {Wang}, {Wang}, {Wang}, {Wang}, {Wang}, {Wang},
  {Wang}, {Wang}, {Wang}, {Wang}, {Wang}, {Xu}, {Xu}, {Yang}, {Yu}, {Yuan},
  {Yuan}, {Zhai}, {Zhang}, {Zhang}, {Zhang}, {Zhao}, {Zhou}, {Zhou}, {Zhu}, \&
  {Zou}}]{2012RAA....12.1197C}
{Cui}, X.-Q., {Zhao}, Y.-H., {Chu}, Y.-Q., {et~al.} 2012, Research in Astronomy
  and Astrophysics, 12, 1197

\bibitem[{{Davenport}(2016)}]{2016ApJ...829...23D}
{Davenport}, J.~R.~A. 2016, \apj, 829, 23

\bibitem[{{De Cat} {et~al.}(2015){De Cat}, {Fu}, {Ren}, {Yang}, {Shi}, {Luo},
  {Yang}, {Wang}, {Zhang}, {Shi}, {Zhang}, {Dong}, {Catanzaro}, {Corbally},
  {Frasca}, {Gray}, {Molenda-{\.Z}akowicz}, {Uytterhoeven}, {Briquet},
  {Bruntt}, {Frandsen}, {Kiss}, {Kurtz}, {Marconi}, {Niemczura}, {{\O}stensen},
  {Ripepi}, {Smalley}, {Southworth}, {Szab{\'o}}, {Telting}, {Karoff}, {Silva
  Aguirre}, {Wu}, {Hou}, {Jin}, \& {Zhou}}]{2015ApJS..220...19D}
{De Cat}, P., {Fu}, J.~N., {Ren}, A.~B., {et~al.} 2015, \apjs, 220, 19

\bibitem[{{Debosscher} {et~al.}(2011){Debosscher}, {Blomme}, {Aerts}, \& {De
  Ridder}}]{2011A&A...529A..89D}
{Debosscher}, J., {Blomme}, J., {Aerts}, C., \& {De Ridder}, J. 2011, \aap,
  529, A89

\bibitem[{{Domingo} {et~al.}(1995){Domingo}, {Fleck}, \&
  {Poland}}]{1995SoPh..162....1D}
{Domingo}, V., {Fleck}, B., \& {Poland}, A.~I. 1995, \solphys, 162, 1

\bibitem[{{Eberhard} \& {Schwarzschild}(1913)}]{1913ApJ....38..292E}
{Eberhard}, G., \& {Schwarzschild}, K. 1913, \apj, 38, 292

\bibitem[{{Fr{\"o}hlich} {et~al.}(1997){Fr{\"o}hlich}, {Crommelynck}, {Wehrli},
  {Anklin}, {Dewitte}, {Fichot}, {Finsterle}, {Jim{\'e}nez}, {Chevalier}, \&
  {Roth}}]{1997SoPh..175..267F}
{Fr{\"o}hlich}, C., {Crommelynck}, D.~A., {Wehrli}, C., {et~al.} 1997,
  \solphys, 175, 267

\bibitem[{{Garc{\'{\i}}a} {et~al.}(2010){Garc{\'{\i}}a}, {Mathur}, {Salabert},
  {Ballot}, {R{\'e}gulo}, {Metcalfe}, \& {Baglin}}]{2010Sci...329.1032G}
{Garc{\'{\i}}a}, R.~A., {Mathur}, S., {Salabert}, D., {et~al.} 2010, Science,
  329, 1032

\bibitem[{{Gilliland} {et~al.}(2015){Gilliland}, {Chaplin}, {Jenkins},
  {Ramsey}, \& {Smith}}]{2015AJ....150..133G}
{Gilliland}, R.~L., {Chaplin}, W.~J., {Jenkins}, J.~M., {Ramsey}, L.~W., \&
  {Smith}, J.~C. 2015, \aj, 150, 133

\bibitem[{{Gilliland} {et~al.}(2011){Gilliland}, {Chaplin}, {Dunham},
  {Argabright}, {Borucki}, {Basri}, {Bryson}, {Buzasi}, {Caldwell}, {Elsworth},
  {Jenkins}, {Koch}, {Kolodziejczak}, {Miglio}, {van Cleve}, {Walkowicz}, \&
  {Welsh}}]{2011ApJS..197....6G}
{Gilliland}, R.~L., {Chaplin}, W.~J., {Dunham}, E.~W., {et~al.} 2011, \apjs,
  197, 6

\bibitem[{{Haas} {et~al.}(2010){Haas}, {Batalha}, {Bryson}, {Caldwell},
  {Dotson}, {Hall}, {Jenkins}, {Klaus}, {Koch}, {Kolodziejczak}, {Middour},
  {Smith}, {Sobeck}, {Stober}, {Thompson}, \& {Van
  Cleve}}]{2010ApJ...713L.115H}
{Haas}, M.~R., {Batalha}, N.~M., {Bryson}, S.~T., {et~al.} 2010, \apjl, 713,
  L115

\bibitem[{{Hall}(2008)}]{2008LRSP....5....2H}
{Hall}, J.~C. 2008, Living Reviews in Solar Physics, 5, 2

\bibitem[{{Hathaway}(2015)}]{2015LRSP...12....4H}
{Hathaway}, D.~H. 2015, Living Reviews in Solar Physics, 12, 4

\bibitem[{{Hawley} {et~al.}(2014){Hawley}, {Davenport}, {Kowalski},
  {Wisniewski}, {Hebb}, {Deitrick}, \& {Hilton}}]{2014ApJ...797..121H}
{Hawley}, S.~L., {Davenport}, J.~R.~A., {Kowalski}, A.~F., {et~al.} 2014, \apj,
  797, 121

\bibitem[{{He} {et~al.}(2008){He}, {Wang}, {Du}, {Li}, {Cui}, {Zhang}, \&
  {He}}]{2008AdSpR..42.1450H}
{He}, H., {Wang}, H., {Du}, Z., {et~al.} 2008, Advances in Space Research, 42,
  1450

\bibitem[{{He} {et~al.}(2014){He}, {Wang}, {Yan}, {Chen}, \&
  {Fang}}]{2014JGRA..119.3286H}
{He}, H., {Wang}, H., {Yan}, Y., {Chen}, P.~F., \& {Fang}, C. 2014, Journal of
  Geophysical Research: Space Physics, 119, 3286

\bibitem[{{He} {et~al.}(2015){He}, {Wang}, \& {Yun}}]{2015ApJS..221...18H}
{He}, H., {Wang}, H., \& {Yun}, D. 2015, \apjs, 221, 18

\bibitem[{{Hempelmann} {et~al.}(2016){Hempelmann}, {Mittag}, {Gonzalez-Perez},
  {Schmitt}, {Schr{\"o}der}, \& {Rauw}}]{2016A&A...586A..14H}
{Hempelmann}, A., {Mittag}, M., {Gonzalez-Perez}, J.~N., {et~al.} 2016, \aap,
  586, A14

\bibitem[{{Jenkins}(2017)}]{KeplerDataProcessingHandbook}
{Jenkins}, J.~M. 2017, \emph{Kepler} Data Processing Handbook (Moffett Field,
  CA: NASA Ames Research Center) KSCI-19081-002

\bibitem[{{Jenkins} {et~al.}(2010{\natexlab{a}}){Jenkins}, {Caldwell},
  {Chandrasekaran}, {Twicken}, {Bryson}, {Quintana}, {Clarke}, {Li}, {Allen},
  {Tenenbaum}, {Wu}, {Klaus}, {Van Cleve}, {Dotson}, {Haas}, {Gilliland},
  {Koch}, \& {Borucki}}]{2010ApJ...713L.120J}
{Jenkins}, J.~M., {Caldwell}, D.~A., {Chandrasekaran}, H., {et~al.}
  2010{\natexlab{a}}, \apjl, 713, L120

\bibitem[{{Jenkins} {et~al.}(2010{\natexlab{b}}){Jenkins}, {Caldwell},
  {Chandrasekaran}, {Twicken}, {Bryson}, {Quintana}, {Clarke}, {Li}, {Allen},
  {Tenenbaum}, {Wu}, {Klaus}, {Middour}, {Cote}, {McCauliff}, {Girouard},
  {Gunter}, {Wohler}, {Sommers}, {Hall}, {Uddin}, {Wu}, {Bhavsar}, {Van Cleve},
  {Pletcher}, {Dotson}, {Haas}, {Gilliland}, {Koch}, \&
  {Borucki}}]{2010ApJ...713L..87J}
---. 2010{\natexlab{b}}, \apjl, 713, L87

\bibitem[{{Karoff} {et~al.}(2016){Karoff}, {Knudsen}, {De Cat}, {Bonanno},
  {Fogtmann-Schulz}, {Fu}, {Frasca}, {Inceoglu}, {Olsen}, {Zhang}, {Hou},
  {Wang}, {Shi}, \& {Zhang}}]{2016NatCo...711058K}
{Karoff}, C., {Knudsen}, M.~F., {De Cat}, P., {et~al.} 2016, Nature
  Communications, 7, 11058

\bibitem[{{Koch} {et~al.}(2010){Koch}, {Borucki}, {Basri}, {Batalha}, {Brown},
  {Caldwell}, {Christensen-Dalsgaard}, {Cochran}, {DeVore}, {Dunham},
  {Gautier}, {Geary}, {Gilliland}, {Gould}, {Jenkins}, {Kondo}, {Latham},
  {Lissauer}, {Marcy}, {Monet}, {Sasselov}, {Boss}, {Brownlee}, {Caldwell},
  {Dupree}, {Howell}, {Kjeldsen}, {Meibom}, {Morrison}, {Owen}, {Reitsema},
  {Tarter}, {Bryson}, {Dotson}, {Gazis}, {Haas}, {Kolodziejczak}, {Rowe}, {Van
  Cleve}, {Allen}, {Chandrasekaran}, {Clarke}, {Li}, {Quintana}, {Tenenbaum},
  {Twicken}, \& {Wu}}]{2010ApJ...713L..79K}
{Koch}, D.~G., {Borucki}, W.~J., {Basri}, G., {et~al.} 2010, \apjl, 713, L79

\bibitem[{{Kowalski} {et~al.}(2013){Kowalski}, {Hawley}, {Wisniewski}, {Osten},
  {Hilton}, {Holtzman}, {Schmidt}, \& {Davenport}}]{2013ApJS..207...15K}
{Kowalski}, A.~F., {Hawley}, S.~L., {Wisniewski}, J.~P., {et~al.} 2013, \apjs,
  207, 15

\bibitem[{{Leka} \& {Barnes}(2003)}]{2003ApJ...595.1296L}
{Leka}, K.~D., \& {Barnes}, G. 2003, \apj, 595, 1296

\bibitem[{{Leka} \& {Barnes}(2007)}]{2007ApJ...656.1173L}
---. 2007, \apj, 656, 1173

\bibitem[{{Maehara} {et~al.}(2012){Maehara}, {Shibayama}, {Notsu}, {Notsu},
  {Nagao}, {Kusaba}, {Honda}, {Nogami}, \& {Shibata}}]{2012Natur.485..478M}
{Maehara}, H., {Shibayama}, T., {Notsu}, S., {et~al.} 2012, \nat, 485, 478

\bibitem[{{McQuillan} {et~al.}(2013){McQuillan}, {Aigrain}, \&
  {Mazeh}}]{2013MNRAS.432.1203M}
{McQuillan}, A., {Aigrain}, S., \& {Mazeh}, T. 2013, \mnras, 432, 1203

\bibitem[{{McQuillan} {et~al.}(2012){McQuillan}, {Aigrain}, \&
  {Roberts}}]{2012A&A...539A.137M}
{McQuillan}, A., {Aigrain}, S., \& {Roberts}, S. 2012, \aap, 539, A137

\bibitem[{{McQuillan} {et~al.}(2014){McQuillan}, {Mazeh}, \&
  {Aigrain}}]{2014ApJS..211...24M}
{McQuillan}, A., {Mazeh}, T., \& {Aigrain}, S. 2014, \apjs, 211, 24

\bibitem[{{Mehrabi} {et~al.}(2017){Mehrabi}, {He}, \&
  {Khosroshahi}}]{2017ApJ...834..207M}
{Mehrabi}, A., {He}, H., \& {Khosroshahi}, H. 2017, \apj, 834, 207

\bibitem[{{Notsu} {et~al.}(2015){Notsu}, {Honda}, {Maehara}, {Notsu},
  {Shibayama}, {Nogami}, \& {Shibata}}]{2015PASJ...67...33N}
{Notsu}, Y., {Honda}, S., {Maehara}, H., {et~al.} 2015, \pasj, 67, 33

\bibitem[{{Notsu} {et~al.}(2013){Notsu}, {Shibayama}, {Maehara}, {Notsu},
  {Nagao}, {Honda}, {Ishii}, {Nogami}, \& {Shibata}}]{2013ApJ...771..127N}
{Notsu}, Y., {Shibayama}, T., {Maehara}, H., {et~al.} 2013, \apj, 771, 127

\bibitem[{{Petrovay}(2010)}]{2010LRSP....7....6P}
{Petrovay}, K. 2010, Living Reviews in Solar Physics, 7, 6

\bibitem[{{Priest} \& {Forbes}(2002)}]{2002A&ARv..10..313P}
{Priest}, E.~R., \& {Forbes}, T.~G. 2002, \aapr, 10, 313

\bibitem[{{Reinhold} {et~al.}(2017){Reinhold}, {Cameron}, \&
  {Gizon}}]{2017A&A...603A..52R}
{Reinhold}, T., {Cameron}, R.~H., \& {Gizon}, L. 2017, \aap, 603, A52

\bibitem[{{Schrijver} {et~al.}(1989){Schrijver}, {Cote}, {Zwaan}, \&
  {Saar}}]{1989ApJ...337..964S}
{Schrijver}, C.~J., {Cote}, J., {Zwaan}, C., \& {Saar}, S.~H. 1989, \apj, 337,
  964

\bibitem[{{Shapiro} {et~al.}(2016){Shapiro}, {Solanki}, {Krivova}, {Yeo}, \&
  {Schmutz}}]{2016A&A...589A..46S}
{Shapiro}, A.~I., {Solanki}, S.~K., {Krivova}, N.~A., {Yeo}, K.~L., \&
  {Schmutz}, W.~K. 2016, \aap, 589, A46

\bibitem[{{Shibata} \& {Magara}(2011)}]{2011LRSP....8....6S}
{Shibata}, K., \& {Magara}, T. 2011, Living Reviews in Solar Physics, 8, 6

\bibitem[{{Shibayama} {et~al.}(2013){Shibayama}, {Maehara}, {Notsu}, {Notsu},
  {Nagao}, {Honda}, {Ishii}, {Nogami}, \& {Shibata}}]{2013ApJS..209....5S}
{Shibayama}, T., {Maehara}, H., {Notsu}, S., {et~al.} 2013, \apjs, 209, 5

\bibitem[{{Skumanich} {et~al.}(1975){Skumanich}, {Smythe}, \&
  {Frazier}}]{1975ApJ...200..747S}
{Skumanich}, A., {Smythe}, C., \& {Frazier}, E.~N. 1975, \apj, 200, 747

\bibitem[{{Smith} {et~al.}(2012){Smith}, {Stumpe}, {Van Cleve}, {Jenkins},
  {Barclay}, {Fanelli}, {Girouard}, {Kolodziejczak}, {McCauliff}, {Morris}, \&
  {Twicken}}]{2012PASP..124.1000S}
{Smith}, J.~C., {Stumpe}, M.~C., {Van Cleve}, J.~E., {et~al.} 2012, \pasp, 124,
  1000

\bibitem[{{Solanki} {et~al.}(2006){Solanki}, {Inhester}, \&
  {Sch{\"u}ssler}}]{2006RPPh...69..563S}
{Solanki}, S.~K., {Inhester}, B., \& {Sch{\"u}ssler}, M. 2006, Reports on
  Progress in Physics, 69, 563

\bibitem[{{Stumpe} {et~al.}(2014){Stumpe}, {Smith}, {Catanzarite}, {Van Cleve},
  {Jenkins}, {Twicken}, \& {Girouard}}]{2014PASP..126..100S}
{Stumpe}, M.~C., {Smith}, J.~C., {Catanzarite}, J.~H., {et~al.} 2014, \pasp,
  126, 100

\bibitem[{{Stumpe} {et~al.}(2012){Stumpe}, {Smith}, {Van Cleve}, {Twicken},
  {Barclay}, {Fanelli}, {Girouard}, {Jenkins}, {Kolodziejczak}, {McCauliff}, \&
  {Morris}}]{2012PASP..124..985S}
{Stumpe}, M.~C., {Smith}, J.~C., {Van Cleve}, J.~E., {et~al.} 2012, \pasp, 124,
  985

\bibitem[{{Su{\'a}rez Mascare{\~n}o} {et~al.}(2015){Su{\'a}rez Mascare{\~n}o},
  {Rebolo}, {Gonz{\'a}lez Hern{\'a}ndez}, \& {Esposito}}]{2015MNRAS.452.2745S}
{Su{\'a}rez Mascare{\~n}o}, A., {Rebolo}, R., {Gonz{\'a}lez Hern{\'a}ndez},
  J.~I., \& {Esposito}, M. 2015, \mnras, 452, 2745

\bibitem[{{Thompson} {et~al.}(2016){Thompson}, {Caldwell}, {Jenkins},
  {Barclay}, {Barentsen}, {Bryson}, {Burke}, {Campbell}, {Catanzarite},
  {Christiansen}, {Clarke}, {Col\'{o}n}, {Cote}, {Coughlin}, {Girouard},
  {Haas}, {Harrison}, {Ibrahim}, {Klaus}, {Li}, {McCauliff}, {Morris},
  {Mullally}, {Rowe}, {Sabale}, {Seader}, {Smith}, {Tenenbaum}, {Twicken},
  {Uddin}, \& J.}]{KeplerDataRelease25}
{Thompson}, S.~E., {Caldwell}, D.~A., {Jenkins}, J.~M., {et~al.} 2016,
  \emph{Kepler} Data Release 25 Notes (Moffett Field, CA: NASA Ames Research
  Center) KSCI-19065-002

\bibitem[{{Van Cleve} {et~al.}(2016){Van Cleve}, {Christiansen}, {Jenkins},
  {Caldwell}, {Barclay}, {Bryson}, {Burke}, {Campbell}, {Catanzarite},
  {Clarke}, {Coughlin}, {Girouard}, {Haas}, {Klaus}, {Kolodziejczak}, {Li},
  {McCauliff}, {Morris}, {Mullally}, {Quintana}, {Rowe}, {Sabale}, {Seader},
  {Smith}, {Still}, {Tenenbaum}, {Thompson}, {Twicken}, {Uddin}, \&
  {Zamudio}}]{KeplerDataHandbook}
{Van Cleve}, J.~E., {Christiansen}, J.~L., {Jenkins}, J.~M., {et~al.} 2016,
  \emph{Kepler} Data Characteristics Handbook (Moffett Field, CA: NASA Ames
  Research Center) KSCI-19040-005

\bibitem[{{Van Doorsselaere} {et~al.}(2017){Van Doorsselaere}, {Shariati}, \&
  {Debosscher}}]{2017ApJS..232...26V}
{Van Doorsselaere}, T., {Shariati}, H., \& {Debosscher}, J. 2017, \apjs, 232,
  26

\bibitem[{{Vaughan} {et~al.}(1981){Vaughan}, {Preston}, {Baliunas}, {Hartmann},
  {Noyes}, {Middelkoop}, \& {Mihalas}}]{1981ApJ...250..276V}
{Vaughan}, A.~H., {Preston}, G.~W., {Baliunas}, S.~L., {et~al.} 1981, \apj,
  250, 276

\bibitem[{{Vaughan} {et~al.}(1978){Vaughan}, {Preston}, \&
  {Wilson}}]{1978PASP...90..267V}
{Vaughan}, A.~H., {Preston}, G.~W., \& {Wilson}, O.~C. 1978, \pasp, 90, 267

\bibitem[{{Walker} {et~al.}(2003){Walker}, {Matthews}, {Kuschnig}, {Johnson},
  {Rucinski}, {Pazder}, {Burley}, {Walker}, {Skaret}, {Zee}, {Grocott},
  {Carroll}, {Sinclair}, {Sturgeon}, \& {Harron}}]{2003PASP..115.1023W}
{Walker}, G., {Matthews}, J., {Kuschnig}, R., {et~al.} 2003, \pasp, 115, 1023

\bibitem[{{Walkowicz} {et~al.}(2011){Walkowicz}, {Basri}, {Batalha},
  {Gilliland}, {Jenkins}, {Borucki}, {Koch}, {Caldwell}, {Dupree}, {Latham},
  {Meibom}, {Howell}, {Brown}, \& {Bryson}}]{2011AJ....141...50W}
{Walkowicz}, L.~M., {Basri}, G., {Batalha}, N., {et~al.} 2011, \aj, 141, 50

\bibitem[{{Wang} {et~al.}(2009){Wang}, {Cui}, \& {He}}]{2009RAA.....9..687W}
{Wang}, H.-N., {Cui}, Y.-M., \& {He}, H. 2009, Research in Astronomy and
  Astrophysics, 9, 687

\bibitem[{{Wang} {et~al.}(1996){Wang}, {Shi}, {Wang}, \&
  {Lue}}]{1996ApJ...456..861W}
{Wang}, J., {Shi}, Z., {Wang}, H., \& {Lue}, Y. 1996, \apj, 456, 861

\bibitem[{{Wang} {et~al.}(2015){Wang}, {Zhang}, {He}, {Chen}, {Jin}, \&
  {Zhou}}]{2015SCPMA..58.5682W}
{Wang}, J., {Zhang}, Y., {He}, H., {et~al.} 2015, Science China Physics,
  Mechanics, and Astronomy, 58, 599601

\bibitem[{{Wilson}(1978)}]{1978ApJ...226..379W}
{Wilson}, O.~C. 1978, \apj, 226, 379

\bibitem[{{Yang} {et~al.}(2014){Yang}, {Chen}, {Hsieh}, {Wu}, {He}, \&
  {Tsai}}]{2014ApJ...786...72Y}
{Yang}, Y.-H., {Chen}, P.~F., {Hsieh}, M.-S., {et~al.} 2014, \apj, 786, 72

\bibitem[{{Yang} {et~al.}(2017){Yang}, {Hsieh}, {Yu}, \&
  {Chen}}]{2017ApJ...834..150Y}
{Yang}, Y.-H., {Hsieh}, M.-S., {Yu}, H.-S., \& {Chen}, P.~F. 2017, \apj, 834,
  150

\bibitem[{{Yun} {et~al.}(2016){Yun}, {Wang}, \& {He}}]{2016AcASn..57....9Y}
{Yun}, D., {Wang}, H.~N., \& {He}, H. 2016, Acta Astronomica Sinica, 57, 9

\bibitem[{{Yun} {et~al.}(2017){Yun}, {Wang}, \& {He}}]{2017ChA&A..41...32D}
---. 2017, \caa, 41, 32

\bibitem[{{Zechmeister} \& {K{\"u}rster}(2009)}]{2009A&A...496..577Z}
{Zechmeister}, M., \& {K{\"u}rster}, M. 2009, \aap, 496, 577

\bibitem[{{Zhang} {et~al.}(2006){Zhang}, {Flyer}, \&
  {Low}}]{2006ApJ...644..575Z}
{Zhang}, M., {Flyer}, N., \& {Low}, B.~C. 2006, \apj, 644, 575

\end{thebibliography}
\end{document}